\documentclass[aps,prb,twocolumn,showpacs,amsmath,amssymb,superscriptaddress,floatfix]{revtex4-2}

\usepackage{blindtext}
\usepackage{latexsym}
\usepackage{amssymb}
\usepackage{physics}
\usepackage{amsmath}
\usepackage{bm}
\usepackage{amsfonts}
\usepackage{relsize}
\usepackage{xcolor}
\usepackage{verbatim}
\usepackage{bbold}
\usepackage{slashed}
\usepackage{appendix}
\usepackage{graphicx}
\usepackage{subfigure}
\usepackage{transparent}
\usepackage[normalem]{ulem}
\usepackage[colorlinks = true,
            linkcolor = blue,
            urlcolor  = blue,
            citecolor = blue,
            anchorcolor = blue]{hyperref}
\usepackage{graphicx} % Include figure files
\usepackage{dcolumn} % Align table columns on decimal point

\newcommand{\SUtwog}{SU(2) }
\newcommand{\msml}[1]{{\mathsmaller{\mathsmaller{\mathsmaller{#1}}}}}
\newcommand{\smsqr}{\msml{\square}}
\newcommand{\bs}[1]{\mathbf{#1}}

\newcommand{\Deltam}[1]{\Delta_{{m}}(\bs{#1})}
\newcommand{\Deltap}[1]{\Delta_{\rm p}(\bs{#1})}

\newcommand{\kk}{\bs{k}}

\newcommand{\qb}{\bs{q}}
\newcommand{\QQ}{\bs{Q}}
\renewcommand{\bar}{\overline}
\newcommand{\Comment}[1]{ }

\newcommand{\KK}{K}
\newcommand{\DDD}{D}

\newcommand{\W}{A}

\newcommand{\Gdue}{\mathbf{G}}
\newcommand{\Fdue}{\mathbf{F}^{p}}

\newcommand{\up}{\uparrow}
\newcommand{\down}{\downarrow}
\newcommand{\Neel}{{N\'eel}}
\newcommand{\DV}[1]{{\color{red}DV: #1}}

\definecolor{ao(english)}{rgb}{0.0, 0.5, 0.0}
\newcommand{\PB}[1]{{\color{ao(english)}PB: #1}}

\begin{document}

% Phase stiffness in an antiferromagnetic superconductor
%\title{Spin stiffness in a superconducting antiferromagnet}
%\title{Magnetic and superconducting order parameter fluctuations in the two-dimensional Hubbard model (change title)}
\title{SC$^*$ Superconductivity and Spin Stiffnesses in the SU(2) Gauge Theory of the Two-Dimensional Hubbard Model}

\author{Demetrio Vilardi}
\affiliation{Max Planck Institute for Solid State Research, D-70569 Stuttgart, Germany}
\author{Pietro M.~Bonetti}
\affiliation{Department of Physics, Harvard University, Cambridge, Massachusetts 02138, USA}
\affiliation{Max Planck Institute for Solid State Research, D-70569 Stuttgart, Germany}
\date{\today}
\begin{abstract}

We consider the SU(2) gauge theory for spin fluctuations in the two-dimensional Hubbard model, where the electron field is fractionalized in terms of spinons and chargons. In this theory, spinons are described by a non-linear sigma model, while chargons are treated as fermions at a mean-field level. We investigate the instability to a superconducting state SC*, arising from a fractionalized Fermi liquid (FL*) where pairing between chargons occurs. Consistent with previous studies, our analysis reveals a coexisting phase characterized by both magnetic and superconducting order for the chargons. The central contribution of this work is the calculation of the feedback of superconductivity on spatial and temporal spin stiffnesses, thereby quantifying its impact on spin fluctuations. Our key finding is that superconductivity significantly suppresses these spin stiffnesses, enhancing quantum spin fluctuations. This enhancement suggests that superconductivity can play a crucial role in stabilizing quantum disorder against long-range magnetic ordering. 

\end{abstract}
\pacs{}
\maketitle

\section{Introduction}

In high-$T_c$ superconductors, the understanding of the pseudogap phase still poses many challenges~\cite{Keimer2015}. 
Many experiments 
%in such phase points towards some sort of short-range 
exhibit features that are compatible with some sort of magnetic order, as, for instance, the rapid change in the Hall number at the transition while entering the phase~\cite{Collignon2017}, or the angle-resolved magnetic resistance experiment associated with Fermi surface reconstruction~\cite{Fang2022}, to name a few. 
However, in many compounds, long range magnetic order extends only to a very few percent of hole-dopings, making the behavior even more puzzling. 

A possible strategy to understand the pseudogap phase has been proposed in Refs.~\cite{Sachdev2016,Chatterjee2017} where the electron field has been fractionalized in terms of a \textit{spinon} and a \textit{chargon} field with the appearing of an emergent \SUtwog gauge symmetry. The spinon field is an SU(2) matrix which rotates the spin reference frame of the chargon field locally in space and time. The chargon field instead represents a spinless fermion, carrying the electromagnetic charge of the electron, as well as an \SUtwog gauge charge. Within such framework, a Higgs field can be introduced~\cite{Chatterjee2017,Scheurer2018,Sachdev2019}, \textit{higgsing down} the gauge group to U(1) or even $\mathbb{Z}_2$, depending on the specific form of the condensate. As long as the spinons remain uncondensed, such Higgs phase represents a topological state with no magnetic order and emerging gauge fields, in other words \textit{a spin liquid}~\cite{SachdevReview2019} coexisting with spinless fermionic degrees of freedom that will form a Fermi surface. The latter is reconstructed by the condensation of the Higgs field, leading to a violation of Luttinger theorem, which is made possible by the emergent fractionalized degrees of freedom~\cite{MO00,FLS1,FLS2}. A quantity that serves as a proxy of the strength of magnetic fluctuations is the \textit{spin stiffness}, a large value of which implies strong tendencies towards long range magnetic ordering, whereas small values enhance quantum fluctuations, thereby leaving the open possibility for topological order to remain stable. 

Moreover, (doped) topological phases can be unstable to other kinds of perturbations, the most relevant of them all being superconductivity. There are essentially two routes to induce superconductivity in a spin liquid. It could occur as a \textit{confining} transition~\cite{Christos23}, that completely higgses out the emergent gauge field, leaving behind a trivial superconducting state that smoothly connects to its analog that one would obtain as an instability of a Fermi liquid~\cite{Christos231}. Alternatively, in our language, it could occur as pairing of \textit{chargons}, realizing a so-called SC$^*$ phase, where superconductivity coexists with fractionalized excitations. Such a superconducting state is separated from a conventional one by a deconfined quantum critical point, describing a nontrivial phase transition. Such SC$^*$ phase has also been studied in the context of Kondo lattice systems in Ref.~\cite{Bunney2025}. 

In this manuscript, we consider an SC$^*$ phase emerging out of the topologically ordered phase of the SU(2) gauge theory for the pseudogap phase described above. While confining superconducting phases are probably more suited to describe the phenomenology of cuprate superconductors~\cite{Christos231}, within the present formalism, simultaneously confining the gauge fields and obtaining Cooper pairing is a nontrivial task. On the other hand, an SC$^*$ state can simply emerge as chargon pairing. A nontrivial question that we aim to address is how the presence of superconductivity influences the spin stiffnesses, parametrizing the low energy dispersion of spinon degrees of freedom and at the same time the strength of quantum fluctuations, which are required to keep the groundstate stable against magnetic order. A useful byproduct of our calculation is that our formulas for the spin stiffnesses also apply to more conventional systems with coexisting magnetic and superconducting orders, as in that case chargons can be interpreted as conventional electrons. States with such coexisting orders have recently been observed in clean samples of cuprate superconductors at low doping~\cite{Kurokawa2023}. A calculation complementary to our work has been performed in Ref.~\cite{Metzner2019}, where the superconducting phase stiffness has been computed in such coexisting order. 

We calculate microscopically the spin stiffnesses starting from the square lattice Hubbard model, which has been proposed as a prototypical minimal model for electrons in the copper-oxide planes. The pseudogap phase and superconductivity have been found in this model, using both numerical calculations~\cite{QinRev2022,Gull2013} and unbiased diagrammatic techniques~\cite{Halboth2000,Vilardi2018,Gunnarsson2015}. 
%The physical understanding of these phases still remains challenging, despite the incredible computational and analytical efforts. 
Similarly to Ref~\cite{Bonetti2022Gauge}, where superconductivity was not considered, we fix the parameters of the gauge theory employing the functional renormalization group (fRG)~\cite{Dupuis2021} in combination with mean-field theory~\cite{Wang2014,Eberlein2016,Vilardi2020}. We find that the presence of superconductivity typically diminishes the value of the spin stiffnesses, thereby enhancing spin fluctuations, which in turn could stabilize topological order with no long range magnetism. This effect on the spin stiffness is expected to persist even in the presence of strong superconducting fluctuations, which exist in the range between the mean-field critical temperature $T_{\rm MF}$ and the Berezinskii-Kosterlitz-Thouless (BKT)~\cite{Berezinskii1971,Kosterlitz_1973} transition temperature $T_{\rm BKT}$.

The paper is structured as follows. In Section~\ref{sec: gauge theory}, we introduce the SU(2) gauge theory and derive an effective low energy model for the low-energy degrees of freedom. In Section~\ref{sec: aux}, we derive formulas for the spin stiffnesses in the SC$^*$ state. In Section~\ref{sec: results} we present our numerical results and, finally, our conclusions are presented in Section~\ref{sec: conclusions}. 
\section{Low energy model}
\label{sec: gauge theory}

\subsection{Spiral and superconducting orders}
\label{sec: mean-field}

Before introducing the low energy theory for spin fluctuations, we briefly discuss the underlying mean-field order considered here, constituted by the coexistence of incommensurate spiral magnetism and superconductivity in a two-dimensional square lattice. 

The spiral order is characterized by a distribution of the average spin operator lying in the $xy$-plane as
\begin{equation}
    \langle \bs{S}_{j} \rangle = m \left[ \cos{}(\bs{Q}\cdot \bs{R}_j)\bs{e}_x + \sin{}(\bs{Q}\cdot \bs{R}_j) \bs{e}_y \right] \, \
\label{eq: spiral}
\end{equation}
where $m$ is the magnetization, $\bs{R}_j$ the position vector of the $j$-th lattice site, $\bs{e}_\alpha$ are unit vectors in real space, with $\alpha=1,2,3$ and $\bs{Q}$ is a fixed wave vector. The choice of the plane for the spiral order is arbitrary, since it is randomly selected once the SU(2) spin symmetry is broken; here, the $xy$-plane is chosen for convenience. We define the spin-charge operator as
\begin{equation}
    S^a_j = \frac{1}{2}\sum_{s,s'} c^\dagger_{j,s} \sigma^a_{ss'} c_{j,s'} \, ,
\label{eq: Sj}
\end{equation}
where $\sigma^a$ ($a=0,1,2,3$) are the Pauli matrices (for $a=1,2,3$), $\sigma^0$ is the 2x2 identity matrix and $c_{j,s}$ ($c^\dagger{j,s}$) is the annihilation (creation) operator in real space. For $a=1,2,3$ Eq.~\eqref{eq: Sj} represents the components of the spin operator appearing in Eq.~\eqref{eq: spiral}, while for $a=0$ it reflects the charge operator. After Fourier transforming Eqs.~\eqref{eq: spiral} and \eqref{eq: Sj}, we obtain the order parameter
\begin{equation}
    m = \int_{\bs{k}}\langle c^\dagger_{\bs{k},\up} c_{\bs{k}+\bs{Q},\down}\rangle \, ,
\end{equation}
where $c_{\bs{k},s}$ and $c^\dagger_{\bs{k},s}$ are creation and annihilation operators in momentum space. Here, we use a short-hand notation for the momentum integration as $\int_{\bs{k}} = \int \frac{d^2\bs{k}}{(2\pi)^2}$. 

The mean-field Hamiltonian for the state with coexisting spiral magnetism and superconductivity takes the form~\cite{Metzner2019}
\begin{align}
\nonumber
    H = &\int_{\kk} \sum_{\sigma} \epsilon_{\kk} c^\dagger_{\kk\sigma}c_{\kk\sigma} \\ &+ 
    \int_\kk \Delta_{{\rm m}} c^\dagger_{\kk+\QQ \downarrow} c_{\kk \uparrow} + 
    \int_\kk \Delta_{{\rm p,} \kk} c^\dagger_{-\kk\downarrow} c^\dagger_{\kk\uparrow} 
    + \mathrm{h.c.} \, ,
\label{eq: Ham}
\end{align}
where $\Delta_{{\rm p},\kk}$ and $\Delta_{\rm m}$ are the magnetic and superconducting gap, respectively. Here, we assume the magnetic gap $\Delta_{\rm m}$ to be momentum independent. 

This state breaks both the spin SU(2) and charge U(1). In the special case where $\QQ = (\pi,\pi)$, i.e. a \Neel \ antiferromagnet, the system has a residual U(1) spin symmetry, referring to the rotations around the magnetization axis. For an incommensurate spiral order with $\QQ = (\pi-2\pi\eta,\pi)$ or $\QQ = (\pi,\pi-2\pi\eta)$, the SU(2) spin group is broken to $\mathbb{Z}_2$. 

\subsection{Non-linear sigma model}

The non-linear sigma model (NL$\sigma$M) is a well-established effective field theory that describes the low-energy fluctuations of an order parameter. It is widely used to capture magnetic fluctuations around ordered spin textures~\cite{Auerbach,Dupuis2002,Bonetti2022Gauge}. Its derivation from microscopic models like the Hubbard Hamiltonian, which forms the basis of our study, has been established in Refs.~\cite{Dupuis2002,Bonetti2022Gauge}. This can be done via fractionalization of the electrons, described in this context by the Grassmann field $c_j$ ($c_j^*$), into
\begin{align*}
    c_j = R_j \psi_j \qquad c^*_j = \psi_j^*R^\dagger_j \, ,
\end{align*}
where $R_j$ is a SU(2) matrix and represents the \emph{spinons}. Such matrix acts on the fermionic spinor field $\psi_{j,\sigma}$ representing the \emph{chargons} and carrying the pseudospin index $\sigma$. From the symmetry point of view, the spinon field $R_j$ carries the SU(2) physical spin while the chargon doesn't transform under spin transformations. The fractionalization introduces an emergent SU(2) gauge symmetry due to the invariance under the new local transformations
\begin{align}
\label{eq: fractionalization}
    \psi_j \rightarrow & V_j\psi_j & \psi^*_j \rightarrow & \psi_j^* V^\dagger_j \\
    R_j \rightarrow & R_j V_j & R^\dagger \rightarrow & V^\dagger_j R^\dagger_j \ . 
\end{align}
The spinon matrix $R_j$ carries the SU(2) gauge charge through its second index while the chargon $\psi_j$ carries the same gauge charge through its spinor index. The physical U(1) charge transformation acts only on the chargon field $\psi_j$ as
\begin{align}
    \psi_j \rightarrow &e^{i\theta_j} \psi_j & \psi^*_j \rightarrow & e^{-i\theta_j} \psi^*_j \ .
\end{align}

The fractionalization in Eqs.~\eqref{eq: fractionalization} can be applied to the Hubbard action, leading to an equivalent formulation of the original problem by the new degrees of freedom, $\psi_j$ and $R_j$, that are treated separately~\cite{Dupuis2002,Bonetti2022Gauge}. In particular, the chargon field can be integrated out via a (renormalized) mean field calculation, leading to an effective action for the spinon field, after a long-wavelength approximation, as
\begin{align}
\label{eq: nlsm}
\nonumber
    S_{\rm NL\sigma M}[\mathcal{R}]& = \int \rm{d}x \Big \{ \frac{1}{2}{\rm Tr} \left[ \mathcal{P}_{\mu\nu}( \partial_{\mu}\mathcal{R}^{\rm T}) \left( \partial_{\nu}\mathcal{R}\right)\right]  
\end{align}
where $\mathcal{R}$ is the adjoint representation of the SU(2) matrix $R_j$
\begin{align}
    \mathcal{R}^{ab}\sigma^b = R^\dagger \sigma^a R
\end{align}
$\sigma^a$ are the Pauli matrices ($a=1,2,3$). The matrix $\mathcal{P}_{\mu\nu}$ can be written in terms of the more commonly known stiffness matrix $\mathcal{J}_{\mu\nu} = \Tr [\mathcal{P}_{\mu\nu}] \mathbb{1} - \mathcal{P}_{\mu\nu}$ and is derived from the integration of the chargon field. All the details can be found in Ref.~\cite{Bonetti2022Gauge}. 

\ifx
Considering the mean-field state introduced in the previous section,  we introduce the NL$\sigma$M as
\begin{align}
\label{eq: nlsm}
\nonumber
    S_{\rm NL\sigma M}[\mathcal{R},\Theta]& = \int \rm{d}x \Big \{ \frac{1}{2}{\rm Tr} \left[ \mathcal{P}_{\mu\nu}( \partial_{\mu}\mathcal{R}^{\rm T}) \left( \partial_{\nu}\mathcal{R}\right)\right] 
    \\& + 
    %J^{s}_{ij}(\partial_i \Theta) (\partial_j \Theta) -\chi_\mathrm{n} (\partial_t \Theta)^2 \\ &
    \mathcal{T}_{\mu\nu}(\partial_\mu \Theta)(\partial_\nu \Theta)
    + (\partial_\mu\Theta) \mathrm{Tr} 
    \big[\mathcal{M}_{\mu\nu} \mathcal{R}^\mathrm{T}\partial_\nu \mathcal{R} 
        \big] \Big\}\,,
\end{align}

Here, the SO(3) matrix $\mathcal{R}(t,\bs{x})$ represents a local rotation for the local spin and the field $\Theta(t,\bs{x})=2\theta(t,\bs{x})$ with $\theta(t,\bs{x})$ being the local phase of the superconducting order parameter. The matrix $\mathcal{P}$ can be written in terms of the more commonly known stiffness matrix $\mathcal{J}_{\mu\nu} = \Tr [\mathcal{P}_{\mu\nu}] \mathbb{1} - \mathcal{P}_{\mu\nu}$. 

The action~\eqref{eq: nlsm} can be derived for purely fermionic systems in different ways. One way is to integrating the fermionic path integral first via Hubbard-Stratonovich transformation and then by expanding the resulting fermion determinant~\cite{Fradkin}, or, alternatively, via explicit fractionalization of the fermion field into chargon and spinon fields, where the spinon acts as a local rotation of the fermion field~\cite{Bonetti2022Gauge}. In both cases, the parameters of the low energy action~\eqref{eq: nlsm} are determined via the formal integration of the fermionic degrees of freedom. 

It is noteworthy commenting the first order coupling term between spin and superconducting fluctuations, represented by the last term in the second line of the action~\eqref{eq: nlsm}. Such coupling term vanishes when computed in the selected mean-field state for the $\psi_j$ field, as a consequence the fields $\mathcal{R}$ and $\Theta$ decouple, each described by an independent NL$\sigma$M. We end up with two uninteracting NL$\sigma$M fields for the groups SU(2) and U(1), which has been separately extensively discussed in the literature. More details on the NL$\sigma$M for the SU(2) field $\mathcal{R}$ can be found in Refs.~\cite{Borejsza2004,Bonetti2022Gauge}. The U(1)-NL$\sigma$M for superconducing fluctuations describes, for insance, the Berezinskii-Kosterlitz-Thouless (BKT) phase at finite temperature in two spatial dimensions.
\fi

The structure of the matrix $\mathcal{P}_{\mu\nu}$ depends on the specific ordering state of the chargon. For instance, the \Neel \ antiferromagnetic order, where the spin texture can be written as $\langle \Vec{S}_j \rangle \sim (-1)^{\bs{r}_j} \bs{e}_x$, has two degenerate Goldstone modes and a residual O(2) symmetry for rotations around the axis defined by $\bs{e}_x$. The stiffness matrix becomes
\begin{align}
    \mathcal{J}_{\mu\nu} = \left( 
    \begin{matrix}
        0 & 0 & 0 \\
        0 & J_{\mu\nu} & 0 \\
        0 & 0 & J_{\mu\nu} \\
    \end{matrix} \right)
\end{align}
with $J_{\mu\nu} = \mathrm{diag}(-Z,J,J)$. In this case, the first term of the action~\eqref{eq: nlsm} simplifies to the known O(3)/O(2)$\simeq S^2$ NL$\sigma$M~\cite{Haldane1983a,Haldane1983b}
\begin{align}
    S_{\rm NL\sigma M}[\bs\Omega]& = \int \mathrm{d} x 
    \left[ Z |\partial_\tau \hat{\bs\Omega}|^2 + J|\nabla\hat{\bs\Omega}|^2 \right]
\end{align}
where the vector field $\hat{\bs{\Omega}}$ has components $\hat{\Omega}^a = \mathcal{R}^{a1}$ and obeys $|\hat{\bs\Omega}|=1$. 

Another state considered here is the spiral antiferromagnet, characterized by three Goldstone modes, one for in-plane and two for out-of-plane fluctuations. In this case the stiffness matrix has the form
\begin{align}
\label{eq: J mat}
    \mathcal{J}_{\mu\nu} = \left( 
    \begin{matrix}
        \frac{1}{2}J^{\perp}_{\mu\nu} & 0 & 0 \\
        0 & \frac{1}{2}J^\perp_{\mu\nu} & 0 \\
        0 & 0 & J^\smsqr_{\mu\nu} \\
    \end{matrix} \right)
\end{align}
with 
\begin{align}
\label{eq: Jout mat}
    J^{\rm \perp}_{\mu\nu} = \left( 
    \begin{matrix}
        -Z^{\rm \perp} & 0 & 0 \\
        0 & J^{\rm \perp}_{xx} & J^{\rm \perp}_{xy} \\
        0 & J^{\rm \perp}_{yx} & J^{\rm \perp}_{yy} \\
    \end{matrix} \right)
\end{align}
for the out-of-plane fluctuations and 
\begin{align}
\label{eq: Jin mat}
    J^{\rm \smsqr}_{\mu\nu} = \left( 
    \begin{matrix}
        -Z^{\rm \smsqr} & 0 & 0 \\
        0 & J^{\rm \smsqr}_{xx} & J^{\rm \smsqr}_{xy} \\
        0 & J^{\rm \smsqr}_{yx} & J^{\rm \smsqr}_{yy} \\
    \end{matrix} \right)
\end{align}
for the in-plane ones.

\ifx
\DV{Just remove the following part if you think we don't need it. }
In this case, the resulting NL$\sigma$M has the symmetry group O(3)$\times$O(2)/O(2) and is characterized by three coupled rotors 
\begin{align}
\nonumber
    S_{\rm NL\sigma M}[\bs\Omega_1,&\bs\Omega_2,\bs\Omega_3,\Theta] = \\ & \sum_i\int \mathrm{d} x 
    \left[ z^i (\partial_\tau \hat{\bs\Omega}_i)^2 + p^i_{\alpha\beta}(\partial_\alpha\hat{\bs\Omega}_i)\cdot (\partial_\beta \hat{\bs{\Omega}_i}) \right]
\end{align}
where the coefficients are $p^1_{\alpha\beta}=p^2_{\alpha\beta} = 1/2J^\smsqr_{\alpha\beta}$, $p^3_{\alpha\beta}=J^\perp_{\alpha\beta} - 1/2J^\smsqr_{\alpha\beta}$, with $\alpha,\beta=1,2,3$, and the same expressions for $z^1=z^2=1/2Z^\smsqr$, $z^3=Z^\perp - 1/2Z^\smsqr$. The rotors $\hat{\bs\Omega}_i$, with components $\hat{\Omega}_i^a = \mathcal{R}^{ai}$, are coupled via the relation $\hat{\bs\Omega}_i \cdot \hat{\bs\Omega}_j = \delta_{ij}$ with $i,j=1,2,3$. 
\fi

\section{Stiffness calculation}
\label{sec: aux}

In this section we compute the stiffness matrix as in Eq.~\eqref{eq: J mat} from the mean-field Hamiltonian~\eqref{eq: Ham}. For the spiral order, this has already been studied in Ref.~\cite{Bonetti2022Ward}, where the stiffnesses have been connected to the zero momentum and frequency limits of the response function to an external SU(2) gauge field. More specifically, as also explained in Refs.\cite{Bonetti2022Ward,*Bonetti2022WardErratum} and~\cite{Katanin2024}, such connection requires the application of a symmetry breaking field $\phi_0$, that needs to be sent to zero only at the end of the calculation. In other words, the limits as the frequency, momentum and the field $\phi_0$ go to zero are highly non trivial, and the correct stiffness is recovered only when the zero field $\phi_0$ limit is performed after the momentum and frequency limits. 

The superfluid stiffness has been computed in Ref.~\cite{Metzner2019} for the coexisting phase, while the magnetic stiffness has not been calculated yet and will be the topic of the next sections.  

%As discussed above, the fermionic field $\psi_j$ might long range order at finite Temperature without violating the Mermin-Wagner theorem for the physical electrons. Here, we consider the case where the group \GaugeGroup is globally broken by coexisting incommensurate spiral-like order in the SU(2) index and $d$-wave pairing of the auxiliary fermions.

\subsection{Gauge response function}

In this section, we derive the formula of the spin stiffness for a generic action $S[\psi,\bar\psi,\W]$ by applying an external local SU(2) gauge field $\W^a_\mu(x)$. We start by defining the generating functional 
\begin{align}
    \mathcal{G}[\W,\Vec{h}] = - \ln \int \mathcal{D}\psi \mathcal{D}\bar\psi\, e^{-S[\psi,\bar\psi,\W] + \int \Vec{h}\cdot \bar{\psi}\Vec{\sigma}\psi}
\label{eq: GW}
\end{align}
By considering functional derivatives, one can compute the system's response to an external SU(2) gauge field. At the lowest order in the gauge field, the induced spin current can be expressed as 
\begin{align}
    j^a_\mu(q) = \int_{q'} K^{h,ab}_{\mu\nu}(q,q') \W^b_\nu(q'),
\end{align}
where we have defined the SU(2) gauge kernel as
\begin{align}
\label{eq: K}
K^{h, ab}_{\mu\nu}(q,q') = -\frac{\delta^{2}G[\W]}{\delta \W^b_\nu(q')\delta \W^a_\mu(-q)}\bigg|_{\W=0}.
\end{align}
The kernel is not computed at vanishing source field, but instead at a finite local symmetry breaking field $\Vec{h}(x)$, whose explicit form depends, in general, on the specific symmetry broken state. In the case of the spiral order, also if coexisting with superconductivity, this field assumes the form $\Vec{h}(x) = h(\cos{\left(\QQ \cdot \bs{x}\right)}, \sin{\left(\QQ \cdot \bs{x}\right)},0)$.

Spiral order breaks also translational symmetry, so that the gauge kernel in~\eqref{eq: K} is expected to be nonzero also for $q\neq q'$, with $q=(\qb,\omega)$ and $q'=(\qb',\omega')$.

The spin stiffness, defined as the static limit of the momentum diagonal component of the SU(2) gauge kernel~\cite{Bonetti2022Ward,Bonetti2022WardErratum,Katanin2024}
\begin{align}
\label{eq: J def}
    J^{ab}_{\alpha\beta} = -\lim_{h \rightarrow 0}\lim_{\qb \rightarrow \bs{0}} K^{h, ab}_{\alpha\beta}(\qb,0),
\end{align}
equals the spatial component of Eq.~\eqref{eq: J def}, i.e. 
\begin{align}
    J^{ab}_{\alpha\beta} = \mathcal{J}^{ab}_{\alpha\beta}.
\label{eq: J K}
\end{align}
Here we defined 
\begin{align}
\label{eq: Kq}
    K^{h, ab}_{\mu\nu}(\qb,\qb,\omega) = \delta_{\qb,\qb} K^{h, ab}_{\mu\nu}(\qb,\omega).
\end{align}
\Comment{
It is possible to prove that, the spin stiffness, given by the spatial component of Eq.~\eqref{eq: J def}, i.e. \PB{Is this not our definition?}
\begin{align}
    J^{ab}_{\alpha\beta} = \mathcal{J}^{ab}_{\alpha\beta}
\label{eq: J K}
\end{align}
equals the static limit of the momentum diagonal component of the SU(2) gauge kernel~\cite{Bonetti2022Ward,Bonetti2022Gauge}
\begin{align}
\label{eq: Jij}
    J^{ab}_{\alpha\beta} = -\lim_{\qb \rightarrow \bs{0}} K^{ab}_{\alpha\beta}(\qb,0),
\end{align}
where we have defined 
\begin{align}
\label{eq: Kq}
    K^{ab}_{\mu\nu}(\qb,\qb,\omega) = \delta_{\qb,\qb} K^{ab}_{\mu\nu}(\qb,\omega).
\end{align}
}
Note that in the above formulas, and from this point on, we have used the convention that the indices labeled as $\mu,\nu$ can be either spatial or temporal, that is, $\mu=(0,x,y)$, whereas $\alpha,\beta$ are only spatial, $\alpha=(x,y)$.

In the low energy theory for magnetic fluctuations, an important role is also played by the \emph{temporal stiffnesses}, given by
\begin{align}
    \chi_\mathrm{dyn}^{ab} = \mathcal{J}^{ab}_{00},
\end{align}
which equals the \emph{dynamical limit} of the gauge kernel~\cite{Bonetti2022Ward,Bonetti2022WardErratum}
\begin{align}
    \label{eq: Z}
    \chi_\mathrm{dyn}^{ab} = \lim_{h\rightarrow 0}\lim_{\omega \rightarrow 0} K^{ab}_{00}(\bs{0},\omega).
\end{align}
Since the magnetic order we are focusing on is of spiral type (with N\'eel order being a limiting case), the spin stiffness matrix $J^{ab}_{\mu\nu}$ can be written explicitly in terms of the out-of-plane and in-plane quantities, $J^{\perp}_{\mu\nu}$ and $J^{\smsqr}_{\mu\nu}$, as given in Eq.\eqref{eq: J mat}. The temporal stiffness takes the following form~\cite{Bonetti2022Ward,Bonetti2022Gauge}
%
%\begin{subequations}
\begin{align}
    \label{eq: chi dyn matrix form}
    &\chi_\mathrm{dyn} = \left(
    \begin{array}{ccc}
        \frac{1}{2}Z^\perp & 0 & 0 \\
        0 & \frac{1}{2}Z^\perp & 0 \\
        0 & 0 & Z^\smsqr
    \end{array}
    \right),
\end{align}
%\end{subequations}
%
where $Z^{\perp}$ and $Z^{\smsqr}$ have been introduced in Eqs.~\eqref{eq: Jout mat} and \eqref{eq: Jin mat}.
\\
\subsection{Mean-field action}

%The path integral of the field $\psi_i$ is performed by a combination of functional renormalization group (fRG) and mean-field theory, as described in Refs.~\cite{Eberlein2016,Vilardi2020}. The high-energy modes are integrated via the fRG by choosing a temperature flow in the static approximation~\cite{Honerkampf,Bonetti2022Gauge}. Regarding this aspect, more details can be found in the Ref.~\cite{Bonetti2022Gauge}. 
%\DV{Should we write how to extract the pairing interaction? I think no. }
%\PB{It's OK to sketch it and refer to the papers}

%The formal integration of high-energy modes will be leading to mean-field like effective action for $\psi_j$, whose functional $G[\W]$ can be then written as
%\begin{align}
%    \mathcal{G}[\W] = - \ln \int \mathcal{D}\psi \mathcal{D}\bar\psi\, e^{-S_{\rm eff}[\psi,\bar\psi,\W]}
%\label{eq: GW eff}
%\end{align}
%where $S_{\rm eff}[\psi,\bar\psi,\W]$ is a quadratic mean-field action for the coexisting phase with spiral magnetic and superconducting orders, as in Ref.~\cite{Eberlein2016,Vilardi2020}. Such action can be written as
We are now ready to explicitly compute the gauge kernel for the mean-field order introduced in Section~\ref{sec: mean-field}. For this scope, we couple the Hamiltonian~\eqref{eq: Ham} to an external SU(2) gauge field, yielding
\begin{align}
\label{eq: S psi}
    S[\psi,\bar\psi,\W] = &S_0[\psi,\bar\psi,\W] + S_{m}[\psi,\bar\psi] 
    +  S_{p}[\psi,\bar\psi,\W] ,
\end{align}
where we have defined
\begin{equation}
\label{eq: expA}
    \begin{split}
        S_0[\psi,\bar\psi,\W] = \int_{\tau}\sum_{jj'}\bar\psi_j\big[&(\partial_\tau-\mu-\W_{0,j})\delta_{jj'}\\
    &+ t_{jj'} e^{-\bs{r}_{jj'}\cdot (\nabla + i \bs{\W}_{j})} \Big]\psi_j,
    \end{split}
\end{equation}
\begin{equation}
\label{eq: expAd}
\begin{split}
    S_{p}[\psi,\bar\psi,\W] = \int_{\tau}\sum_{jj'}
    \Delta_{{p},jj'}\, \bar\psi_j &\left[ i\sigma^2 e^{-\bs{r}_{jj'}\cdot (\nabla + i \bs{\W}_{j})}\right]\bar\psi_j 
    %\\ &
    +\mathrm{c.c.}
\end{split}
\end{equation}
and
\begin{align}
    \label{eq: magnetic action}
    &S_{m}[\psi,\bar\psi] = \int_k \Delta_{ m} [ \bar\psi_{k+Q\down} \psi_{k\up} + \bar\psi_{k\up}\psi_{k+Q\down} ] . 
\end{align}
\ifx
The pairing two-body term is divided into singlet and triplet component 
\begin{equation}
    S_{p}[\psi,\bar\psi] =
    \int_k [ \Deltap{k} \bar\psi_{-k\down}\bar\psi_{k\up} + \Delta_{\rm p}^*(\bs{k}) \psi_{k\up}\psi_{-k\down} ]
\label{eq: S sc}
\end{equation}
\begin{align}
    S^t_{p}[\psi,\bar\psi] &=
    \int_k \Delta_{t}^{\up}(\bs{k}) \bar\psi_{k\up}\bar\psi_{-k-Q\up} + \Delta_{t}^{\up*}(\bs{k}) \psi_{-k-Q\up}\psi_{k\up} ] \\&+
    \int_k \Delta_{t}^{\down}(\bs{k}) \bar\psi_{k+Q\down}\bar\psi_{-k\down} + \Delta_{t}^{\down *}(\bs{k}) \psi_{-k\down}\psi_{k+Q\down} ]
\end{align}
\fi

We stress that also the superconducting gap $\Delta_{p,jj'}$ couples to the gauge field $\W$. In fact, since the former is non-local in space, coupling the system with an external gauge field produces a change in the phase when going from the lattice point $j$ to $j'$~\cite{Verma2021}. In other words, the coupling~\eqref{eq: expAd} ensures the invariance of the action under the gauge transformation 
\begin{align}
    \psi_j &\rightarrow \psi^{'} = \mathcal{V}_j\psi_j \\ 
    \W_j &\rightarrow \W^{'}_{\mu,j} = \mathcal{V}^\dagger_j \W_{\mu,j} \mathcal{V}_j + i\mathcal{V}^\dagger_j \partial_\mu \mathcal{V}_j \ . 
\end{align}

\ifx
As discussed in the previous section, we neglect the triplet superconducting state and limit the coexistence to spiral magnetism and singlet pairing. Indeed, we consider the effective interaction only in the charge, magnetic and singlet pairing channels. In such case, the formula for the susceptibility, Eq.~\eqref{eq: tilde chi} simplifies. In fact, by neglecting the triplet channel, the effective interaction becomes a 6x6 matrix in the basis $\{\Gamma^a, \Phi^0, A^0\}$ 
\begin{widetext}
\begin{align}
\label{eq: Interaction}
    \widetilde{V}(q) = \left(\begin{matrix}
        -U_c(q) & 0 & 0 & 0 & 0 & 0 \\
         0 & U_m(q-Q) & 0 & 0 & 0 & 0 \\
         0  & 0 & U_m(q-Q) & 0 & 0 & 0 \\
         0 & 0 & 0 & U_m(q) & 0 & 0 \\
         0 & 0 & 0 & 0 & U_p(q) & 0 \\
         0 & 0 & 0 & 0 & 0 & U_p(q) 
    \end{matrix}
    \right)
\end{align}
\end{widetext} 
\fi

\subsection{Evaluation of the spatial spin stiffness}

We are now ready to compute the spin stiffness by using the formula~\eqref{eq: J K}, together with the functional differentiation, as in Eq.~\eqref{eq: K}, with the action in Eq.~\eqref{eq: S psi}. 

We expand the exponential in Eqs.~\eqref{eq: expA} and~\eqref{eq: expAd} to second order in the gauge field $\W_j$, giving rise to
\begin{equation}\label{eq: S0A}
\begin{split}
    S_0&[\psi,\bar\psi,A] = \frac{1}{2} \int_{k,q} A^a_\mu (q)\,
    \gamma^\mu_{\bs{k}}\, \bar\psi_{k+q}\sigma^a \psi_k 
    \\ 
    &- \frac{1}{8} \int_{k, q, q'} A^a_\mu(q-q') A^a_\nu(q')\, \gamma^{\mu\nu}_{\bs{k}} \,\bar\psi_{k+q} \psi_{k},
\end{split}
\end{equation}
and, similarly, 
\begin{equation}\label{eq: Sp[A]}
\begin{split}
    S_{p}&[\psi,\bar\psi,A] = \frac{1}{2} \int_{k,q} A^a_\mu (q) 
    \Delta_{{p},\bs{k}}^\mu\, \bar\psi_{k+q}(i\sigma^2\sigma^a) \bar\psi_{-k}
    \\ &- \frac{1}{8} \int_{k, q, q'} A^a_\mu(q-q') A^a_\nu(q') \Delta_{{p},\bs{k}}^{\mu\nu}\,\bar\psi_{k+q}(i\sigma^2) \bar\psi_{-k} 
    %\\ & 
    + \mathrm{c.c.} \hskip 1mm.
\end{split}
\end{equation}
Here, we have also defined
\begin{subequations}
\begin{align}
    &\gamma^{\mu}_{\bs{k}} = (1 \,|\, \partial_{k_\alpha}\xi_{\bs{k}}), \\
    &\gamma^{\mu\nu}_{\bs{k}} = \left(\begin{array}{c|c}0&0 \\ \hline 0&\partial_{k_\alpha}\partial_{k_\beta}\xi_{\bs{k}}\end{array}\right), \\
    & \Delta^{\mu}_{\bs{k}} = (0\,|\,\partial_{k_\alpha}\Delta_{{p}, \bs{k}}), \\
    & \Delta^{\mu\nu}_{\bs{k}} = \left(\begin{array}{c|c}0&0 \\ \hline 0&\partial_{k_\alpha}\partial_{k_\beta}\Delta_{{p}, \bs{k}}\end{array}\right),
\end{align}
\end{subequations}
with $\alpha=x,y$, and $\partial_{k_\alpha}$ a shortcut for $\partial/\partial k_\alpha$.

It is convenient to introduce a 4-component spinor $\Psi_k = \left( \psi_{k\up}, \psi_{k\down}, \bar\psi_{-k\up}, \bar\psi_{-k\down} \right)^\mathrm{T}$. In this basis, the full action can be rewritten as
\begin{equation}
\begin{split}
    S[\psi,&\bar\psi, A] = \int_{k,k'} \bar\Psi_k V_{kk'}[A] \Psi_{k'} 
    \\& + \frac{1}{16T} \int_{q,\bs{k}} \gamma^{\mu\nu}_{\bs{k}} A^{a}_{\mu}(-q) A^{a}_{\nu}(q) + \frac{1}{2T}\int_{\kk}\xi_\kk \ .
\end{split}
\end{equation}
where $V_{kk'}[A]$ reads
\\[2pt]
\begin{equation}
\begin{split}
    V_{kk'}[A] =& -\frac{1}{2}\mathcal{G}^{-1}_{\bs{k}\bs{k}'}(\nu)\delta_{\nu,\nu'} + \frac{1}{4} A^a_\mu(k-k') P^{\mu,a}_{\bs{k}\bs{k'}} \\- &\frac{1}{16} \int_q A^a_\mu(k-k'-q)A^a_\nu (q) P^{\mu\nu}_{\bs{k}\bs{k}'} \ .
\end{split}
\end{equation}
Here we defined $P^{\mu,a}_{\kk\kk'}=\Gamma^{\mu,a}_{\bs{k}\bs{k}'} + \DDD^{\mu,a}_{\bs{k}\bs{k}'}$, $P^{\mu\nu}_{\kk\kk'}=\Gamma^{\mu\nu}_{\bs{k}\bs{k}'} + \DDD^{\mu\nu}_{\bs{k}\bs{k}'}$, with
\begin{subequations}
\begin{align}
\label{eq: gamma1}
    \Gamma^{\mu,a}_{\bs{k}\bs{k}'} &= \left(\begin{matrix}
        \gamma^{\mu}_{\bs{k}'} \sigma^a & 0 \\
        0 & \gamma^{\mu}_{\bs{k}} \left(\sigma^a\right)^{\mathrm{T}} \end{matrix}\right),\\
    \Gamma^{\mu\nu}_{\bs{k}\bs{k}'} &= \left(\begin{matrix}
        \gamma^{\mu\nu}_{\bs{k}'} \mathbb{1} & 0 \\
        0 & - \gamma^{\mu\nu}_{\bs{k}} \mathbb{1} \end{matrix}\right),
\end{align}
\end{subequations}
and
\begin{subequations}
\begin{align}
\label{eq: d1}
    &\DDD^{\mu,a}_{\bs{k}\bs{k}'} = \left(\begin{matrix}
        0 & t^a \Delta_{\bs{k}'}^{\mu} \\
        t^{a\dagger} \Delta_{\bs{k}}^{\mu} & 0 \end{matrix}\right),\\
    &\DDD^{\mu\nu}_{\bs{k}\bs{k}'} = \left(\begin{matrix}
        0 & i\sigma^2 \Delta_{\bs{k}'}^{\mu\nu} \\
        -i\sigma^2 \Delta_{\bs{k}'}^{\mu\nu} & 0 \end{matrix}\right).
\end{align}
\end{subequations}
In the formulas above, $\mathcal{G}_{\bs{k}\bs{k}'}(\nu)$ is the Green's function, which, in this basis, can be written as
\begin{widetext}
\ifx
\begin{align*}
    \mathcal{G}^{-1}_{\bs{k}\bs{k}'}(\nu) =  
    \left(\begin{matrix}
        [i\nu - \xi_{\bs{k}}]\delta_{\bs{k},\bs{k}'} && -\Deltam{k} \delta_{\bs{k+Q},\bs{k}'} && -\Delta_{t}^{\up}(\bs{k})\delta_{\bs{k}+\bs{Q},\bs{k}'} && \Deltap{k}\delta_{\bs{k},\bs{k}'}  \\
        -\Deltam{k'}\delta_{\bs{k-Q},\bs{k}'} && [i\nu - \xi_{\bs{k}}]\delta_{\bs{k},\bs{k}'}  && -\Delta_{\rm p}(-\bs{k})\delta_{\bs{k},\bs{k}'} && -\Delta_{t}^{\down}(\bs{k}-\bs{Q})\delta_{\bs{k}-\bs{Q},\bs{k}'} \\
         -\Delta_{t}^{\up*}(\bs{k}')\delta_{\bs{k}-\bs{Q},\bs{k}'} && -\Delta^*_{\rm p}(-\bs{k})\delta_{\bs{k},\bs{k}'} && [i\nu + \xi_{-\bs{k}}]\delta_{\bs{k},\bs{k}'}  && \Deltam{-k'}\delta_{\bs{k}',\bs{k+Q}} \\
        \Delta_{\rm p}^*(\bs{k})\delta_{\bs{k},\bs{k}'} && -\Delta_{t}^{\down *}(\bs{k})\delta_{\bs{k}+\bs{Q},\bs{k}'} && \Deltam{-k}\delta_{\bs{k}',\bs{k}-\bs{Q}}&& [i\nu + \xi_{-\bs{k}}]\delta_{\bs{k},\bs{k}'}
    \end{matrix}\right),
\end{align*}
\fi
\begin{align}
    \label{eq: inverse Gkkp}
    \mathcal{G}^{-1}_{\bs{k}\bs{k}'}(\nu) =  
    \left(\begin{matrix}
        (i\nu - \xi_{\bs{k}})\delta_{\bs{k},\bs{k}'} && -\Delta_{m} \delta_{\bs{k+Q},\bs{k}'} && 0 && -\Delta_{{p},\bs{k}}\,\delta_{\bs{k},\bs{k}'}  \\
        -\Delta_{m}\,\delta_{\bs{k-Q},\bs{k}'} && (i\nu - \xi_{\bs{k}})\delta_{\bs{k},\bs{k}'}  && \Delta_{{p},\bs{k}}\,\delta_{\bs{k},\bs{k}'} && 0 \\
         0 && \Delta^*_{{p},\bs{k}}\,\delta_{\bs{k},\bs{k}'} && (i\nu + \xi_{\bs{k}})\delta_{\bs{k},\bs{k}'}  && \Delta_{m}\,\delta_{\bs{k}',\bs{k+Q}} \\
        -\Delta^*_{{p},\bs{k}}\,\delta_{\bs{k},\bs{k}'} && 0 && \Delta_{m}\,\delta_{\bs{k}',\bs{k}-\bs{Q}}&& (i\nu + \xi_{\bs{k}})\delta_{\bs{k},\bs{k}'}
    \end{matrix}\right).
\end{align}
\end{widetext}

The full $\mathcal{G}_{\bs{k}\bs{k}'}(\nu)$ is then obtained inverting Eq.~\eqref{eq: inverse Gkkp} both in the spinor and momentum indices. 

We now evaluate the gauge kernel in Eq.~\eqref{eq: K}. Inspecting the structure of the actions~\eqref{eq: S0A} and \eqref{eq: Sp[A]}, one can deduce that it consists of a paramagnetic and a diamagnetic term
\begin{equation}
\begin{split}
    K_{\mu\nu}^{ab}(\bs{q},\bs{q}',\omega) = K_{\mu\nu}^{{\rm p},ab}&(\bs{q},\bs{q}',\omega) 
    \\&+ \delta_{ab} \delta_{\bs{q},\bs{0}} \delta_{\bs{q}',\bs{0}} \delta_{\omega,0}K_{\mu\nu}^{{\rm d}} . 
\end{split}
\end{equation}
The diamagnetic contribution reads as
\begin{align}
    K_{\mu\nu}^{{\rm d}} = -\frac{1}{8}
    \int_{\bs{k},\bs{k}'}T\sum_{\nu_n} \Tr \left[P^{\mu\nu}_{\bs{k}'\bs{k}} 
    \mathcal{G}_{\bs{k}\bs{k}'}(\nu_n) \right].
\end{align}
\ifx
and it can be split in two terms $K_{\mu\nu}^{\rm d} = \KK_{\mu\nu}^{{\rm d}, \Gamma} + \KK_{\mu\nu}^{{\rm d},D}$, with
\begin{align}
    \KK_{\mu\nu}^{{\rm d},\Gamma} &= -\frac{1}{16}
    \int_{\bs{k},\bs{k}'}T\sum_{\nu_n} \Tr \left[\Gamma^{\mu\nu}_{\bs{k}\bs{k}'} 
    \mathcal{G}_{\bs{k},\bs{k}'}(\nu_n) \right] \\
    &= - \frac{1}{8}
    \int_{\bs{k}} T\sum_{\nu_n} \gamma^{\mu\nu}_{\bs{k}} \left[G_{\bs{k}}(\nu) + \bar{G}_{\bs{k}-\bs{Q}}(\nu) \right], 
\end{align}
and
\begin{align}
    \KK_{\mu\nu}^{{\rm d},D} &=
    -\frac{1}{16}
    \int_{\bs{k},\bs{k}'}T\sum_{\nu_n} \Tr \left[\DDD^{\mu\nu}_{\bs{k}\bs{k}'} 
    \mathcal{G}_{\bs{k}\bs{k}'}(\nu_n) \right]
    \\ 
    & = - \frac{1}{8}
    \int_{\bs{k}} T\sum_{\nu_n} \Delta^{\mu\nu}_{\bs{k}} \left[F^s_{\bs{k}}(\nu) + F^s_{-\bs{k}}(-\nu) \right]. 
\end{align}
\fi
The paramagnetic contribution is given by
\begin{equation} \label{eq: param kernel}
\begin{split}
    K_{\mu\nu}^{{\rm p},ab}&(\bs{q},\bs{q}',\omega) = -\frac{1}{4}
    \int_{\bs{k}\bs{k}'}T\sum_{\nu_n} \Tr \Big[P^{\mu,a}_{\bs{k}+\bs{q},\bs{k}} \\ & 
    \times \mathcal{G}_{\bs{k}\bs{k}'}(\nu_n) P^{\nu,b}_{\bs{k}',\bs{k}'+\bs{q}'} \mathcal{G}_{\bs{k}'+\bs{q}',\bs{k}+\bs{q}}(\nu_n+\Omega_m)\Big],
\end{split}
\end{equation}
with $i\Omega_m \rightarrow \omega + i0^+$. 

In order to gain further insight, it is useful to cast the $4\times 4$ Green's function as a block matrix
\begin{align}
\label{eq: Gf}
    \mathcal{G}_{\bs{k}\bs{k}'}(\nu_n) = 
    \left( \begin{matrix}
        \Gdue_{\bs{k}\bs{k}'}(\nu_n)   &   \Fdue_{\bs{k}\bs{k}'}(\nu_n) \\
        \left[\Fdue_{\bs{k}'\bs{k}}(\nu_n)\right]^\dagger  &  -\Gdue^{\mathrm{T}}_{-\bs{k}',-\bs{k}}(-\nu_n)
    \end{matrix}\right),
\end{align}
where $\Gdue_{\bs{k}\bs{k}'}(\nu_n)$ and $\Fdue_{\bs{k}\bs{k}'}(\nu_n)$ are $2\times2$ matrices, representing the spin resolved normal and anomalous propagators, respectively. The normal propagator is parametrized as 
%are 2x2 matrices with components defined as  
%$[\Gdue_{kk'}]_{\sigma\sigma'}\sim\langle \psi_{k'\sigma'} \bar\psi_{k\sigma}\rangle$ and $\left(\Fdue_{k,k'}\right)_{\sigma\sigma'}\sim\langle \bar\psi_{-k'\sigma'}\bar\psi_{k\sigma} \rangle$.
%
\begin{align}
\label{eq: Gdue}
    \Gdue_{\bs{k}\bs{k}'}(\nu_n) = \left(\begin{matrix}
      G_{\bs{k}}(\nu_n)\,\delta_{\bs{k},\bs{k}'}  &  F^{{m}}_{\bs{k}}(\nu_n)\,\delta_{\bs{k},\bs{k}'-\bs{Q}} \\
      F^{{{m}}}_{\bs{k}-\bs{Q}}(\nu_n)\delta_{\bs{k},\bs{k}'+\bs{Q}} & \bar{G}_{\bs{k}-\bs{Q}}(\nu_n)\,\delta_{\bs{k},\bs{k}'}
    \end{matrix}\right).
\end{align}
The anomalous one reads as 
\ifx
(\DV{Still don't know where the relation between $F^{t\up}$ and $F^{t\down}$ comes from. From python I saw $\tilde{F}^{t\down}_\kk = \left[\tilde{F}^{t\up}_\kk\right]^*$, which in the unrotated basis translates to $F^{t\down}_\kk = \left[F^{t\up}_{\kk-\QQ}\right]^*$})
(OLD:)
\begin{align}
\label{eq: Fdue}
    \Fdue_{\bs{k}\bs{k}'}(\nu_n) = \left(\begin{matrix}
      F^{{t}\up}_{\bs{k}}(\nu_n)\delta_{\bs{k}+\bs{Q},\bs{k}'}&  F^{s}_{\bs{k}}(\nu_n)\delta_{\bs{k},\bs{k}'} \\
      -F^{s}_{-\bs{k}'}(-\nu_n)\delta_{\bs{k},\bs{k}'} & F^{{t}\down}_{\bs{k}}(\nu_n)\delta_{\bs{k},\bs{k}'+\bs{Q}}
    \end{matrix}\right),
\end{align}
(NEW: with $\Delta_p\in \mathbb{R}$)
\begin{align}
\label{eq: Fdue}
    \Fdue_{\bs{k}\bs{k}'}(\nu_n) = \left(\begin{matrix}
      F^{t}_{\bs{k}}(\nu_n)\delta_{\bs{k}+\bs{Q},\bs{k}'}&  F^{s}_{\bs{k}}(\nu_n)\delta_{\bs{k},\bs{k}'} \\
      -F^{s}_{-\bs{k}'}(-\nu_n)\delta_{\bs{k},\bs{k}'} & F^{t*}_{\bs{k}-\QQ}(\nu_n)\delta_{\bs{k},\bs{k}'+\bs{Q}}
    \end{matrix}\right),
\label{eq: Fmat}
\end{align}
NEW without $\Delta_p \in \mathbb{R}$
\fi
\begin{align}\label{eq: Fmat}
%\label{eq: Fdue}
    \Fdue_{\bs{k}\bs{k}'}(\nu_n) = \left(\begin{matrix}
      F^{t}_{\bs{k}}(\nu_n)\delta_{\bs{k}+\bs{Q},\bs{k}'}&  F^{s}_{\bs{k}}(\nu_n)\delta_{\bs{k},\bs{k}'} \\
      -F^{s}_{-\bs{k}'}(-\nu_n)\delta_{\bs{k},\bs{k}'} & -F^{t}_{-\bs{k}}(-\nu_n)\delta_{\bs{k},\bs{k}'+\bs{Q}}
    \end{matrix}\right),
\end{align}
and obeys the relation $\Fdue_{\bs{k},\bs{k}'}(\nu_n) = -[\Fdue_{-\bs{k}',-\bs{k}}(-\nu_n)]^\mathrm{T}$. Note that the specific form of the matrix Eq.~\eqref{eq: Fmat} relies on the absence of a spin-triplet superconducting gap.

In general, after selecting the $\bs{q}=\bs{q}'$ component in Eq.~\eqref{eq: param kernel}, and taking the $\bs{q}\to\bs{0}$ limit, after having set $\omega=0$, one obtains expressions for the spin stiffnesses. Decomposing the paramagnetic current vertex as $P_{\bs{k}\bs{k}'}^{\mu,a}=\Gamma_{\bs{k}\bs{k}'}^{\mu,a}+D_{\bs{k}\bs{k}'}^{\mu,a}$ (cf. Eqs.~\eqref{eq: gamma1} and \eqref{eq: d1}), we have
\begin{equation} \label{eq: stiffness decomposition}
    J^X_{\alpha\beta} = J^{\rm d}_{\alpha\beta} + J^{X,\Gamma\Gamma}_{\alpha\beta} + J^{X,DD}_{\alpha\beta}
    + J^{X,\Gamma D}_{\alpha\beta} + J^{X,D\Gamma}_{\alpha\beta},
\end{equation}
with $X=\perp,\smsqr$. The first term in the equation above represents the diamagnetic contribution to the spin stiffness and it is the same for the in- and out-of-plane components. The $J^{X,\Gamma\Gamma}_{\alpha\beta}$ and $J^{X,DD}_{\alpha\beta}$ terms are obtained by setting $P^{a,\mu}_{\bs{k}\bs{k}'}=\Gamma^{a,\mu}_{\bs{k}\bs{k}'}$ and $P^{a,\mu}_{\bs{k}\bs{k}'}=D^{a,\mu}_{\bs{k}\bs{k}'}$ in Eq.~\eqref{eq: param kernel}, respectively. Finally, in the term $J^{X,\Gamma D}_{\alpha\beta}$ ($J^{X,D\Gamma}_{\alpha\beta}$) we have set one the first (second) current vertex in Eq.~\eqref{eq: param kernel} equal to $\Gamma^{a,\mu}_{\bs{k}\bs{k}'}$ and the other one $D^{a,\mu}_{\bs{k}\bs{k}'}$. Note that it can be shown that $J^{X,\Gamma D}_{\alpha\beta}=J^{X,D\Gamma}_{\beta\alpha}$, so that only the evaluation of one of the last two terms in Eq.~\eqref{eq: stiffness decomposition} is required. 
%
%In the case of zero superconducting order parameter ($\Delta_{{p},\bs{k}}=0$), the equations above simplify and only the terms $J^\mathrm{d}_{\alpha\beta}$ and $J^{\smsqr,\Gamma\Gamma}_{\alpha\beta}$ remain finite. We have checked that in this limit we recover the formulas of Ref.~\cite{Bonetti2022Gauge}, obtained in presence of the sole magnetic order. 
%
\begin{widetext}
In Appendix~\ref{app: param and diam terms}, we show the explicit expressions for each of the terms in Eq.~\eqref{eq: stiffness decomposition}. Integrating the diamagnetic term by parts and summing all terms together, we find for the in-plane stiffness
\begin{equation}
\label{eq: J in}
\begin{split}
    J^{\smsqr}_{\alpha\beta} = 
    \int_{\bs{k}} T\sum_{\nu_n} \Big\{
    &-\gamma^\alpha_{\bs{k}}\gamma^\beta_{\bs{k}+\bs{Q}}\left( [F^{{m}}_{\bs{k}}(\nu_n)]^2 -[F^{t}_{\bs{k}}(\nu_n)]^2\right)
    -\Delta^\alpha_{\bs{k}}\Delta^\beta_{\bs{k}+\bs{Q}}\left( |F^{{{m}}}_{\bs{k}}(\nu_n)|^2-|F^t_\bs{k}(\nu_n)|^2\right)\\
    &+2(\gamma^\alpha_{\bs{k}}\Delta^{\beta}_{\bs{k}+\bs{Q}} + \gamma^\beta_{\bs{k}}\Delta^{\alpha}_{\bs{k}+\bs{Q}} )F^m_\bs{k}(\nu_n)F^t_\bs{k}(\nu_n) 
    \Big\},
\end{split}
\end{equation}
and for the out-of-plane one
\begin{equation}
\label{eq: J out}
\begin{split}
    J^{\perp}_{\alpha\beta} =  \frac{1}{2} J^\smsqr_{\alpha\beta}+\frac{1}{2}\int_{\bs{k}} T\sum_{\nu_n} \Big\{
    &\gamma^\alpha_{\bs{k}}\gamma^\beta_{\bs{k}}\left[
    G_{\bs{k}}(\nu_n)\left(G_{-\bs{k}}(\nu_n)-G_{\bs{k}}(\nu_n) \right)
    -F^s_{\bs{k}}(\nu_n)\left(F^s_{-\bs{k}}(\nu_n)-F^s_{\bs{k}}(\nu_n)\right)\right] \\
    &-\Delta^\alpha_{\bs{k}}\Delta^\beta_{\bs{k}}\left[
    G_{\bs{k}}(\nu_n)\left(G^*_{-\bs{k}}(\nu_n)-G^*_{\bs{k}}(\nu_n) \right)
    -F^s_{\bs{k}}(\nu_n)\left(F^s_{-\bs{k}}(\nu_n)-F^s_{\bs{k}}(\nu_n)\right)\right] \\
    &+2(\gamma^\alpha_{\bs{k}}\Delta^{\beta}_{\bs{k}}+\gamma^\beta_{\bs{k}}\Delta^{\alpha}_{\bs{k}})G_{\bs{k}}(\nu_n)\left(F^s_{-\bs{k}}(\nu_n)-F^s_{\bs{k}}(\nu_n)\right)\Big\}.
\end{split}
\end{equation}
\end{widetext}
It is possible to analytically perform the Matsubara summation, as shown in the Appendix~\ref{app: Matsubara}. 

Here, we simplify the momentum dependence of the gap function via a simple form-factor projection: $s$-wave sector for the magnetic gap and $d$-wave for the superconducting gap.  
\subsection{Temporal stiffness}

In this section,  we compute the temporal stiffness, defined as the dynamical limit of the temporal component of gauge response Kernel (see Eq.~\eqref{eq: Z}). Analyzing the coupling of the electrons to the gauge field (Eqs.~\eqref{eq: S0A} and \eqref{eq: Sp[A]}), one can convince himself that the $\mu=\nu=0$ components of the gauge Kernel correspond to the spin susceptibilities.
\begin{align}
    K^{ab}_{00}(\qb,\qb',\omega) = \chi^{ab}(\qb,\qb',\omega) \sim \langle S^a_{-\qb,-\omega}S^b_{\qb',\omega}\rangle\,,
\end{align}
with $S^a_{q=(\qb,\omega)} = (1/2)\int_k \bar{\psi}_{k+q} \sigma^a \psi_k$ being the spin bilinear. As discussed in Refs.~\cite{Kampf1996,Bonetti2022Landau}, it is convenient to work in a rotated spin reference frame in which translational invariance is restored. Such basis is defined by rotating the fermion field $\psi_i$ by an angle $\theta_j=\bs{Q}\cdot \bs{R}_j$ in the $xy$ plane
\begin{align}
    \widetilde\psi_j = e^{-i\frac{\theta_j}{2}} e^{i\frac{\sigma^3}{2}\theta_j}\psi_j.
\end{align}
In this basis, the real-time retarded susceptibility is defined as
\begin{align}
    \widetilde\chi^{ab}(j-j',t) = -i\Theta(t)\left\langle \left[\widetilde{S}^a_j(t), \widetilde{S}^b_{j'}(0)\right] \right\rangle,
\end{align}
with 
\begin{align}
    \widetilde{S}^a_j(t) = \bar{\widetilde{\psi}}_j(t) \frac{\sigma^a}{2} \widetilde{\psi}_j(t).
\end{align}
The susceptibilities in the physical reference frame are obtained from the matrix relation
\begin{align}
    \chi(j,j',t) = M_j^\mathrm{T} \widetilde{\chi}(j-j',t) M_{j'}.
\end{align}
The transformation matrix is given by
\begin{align}
    M_j = \left(
    \begin{matrix}
    \cos(\bs{Q}\cdot \bs{R}_j) & \sin(\bs{Q}\cdot \bs{R}_j) & 0 \\
    -\sin(\bs{Q}\cdot \bs{R}_j) & \cos(\bs{Q}\cdot \bs{R}_j) & 0 \\
    0 & 0 & 1 
    \end{matrix}
    \right).
\end{align}
In Fourier space, the momentum- and spin-diagonal physical susceptibilities read as
\begin{subequations}
\begin{align}
    %\chi^{00}(\qb,\qb,\omega) &= \widetilde{\chi}^{00}(\qb,\omega) \nonumber \\
    \label{eq: chi22 orig to rot}
    \chi^{11}(\bs{q},\bs{q},\omega) &= \chi^{22}(\bs{q},\bs{q},\omega), \\ \nonumber &=
    \widetilde{\chi}^{-+}(\bs{q}+\bs{Q},\omega) + \widetilde{\chi}^{+-}(\bs{q}-\bs{Q},\omega), \\
    \chi^{33}(\qb,\qb,\omega) &= \widetilde{\chi}^{33}(\qb,\omega).
\end{align}
\end{subequations}
with $\widetilde{S}^{\pm}_q = \left( \widetilde{S}^1_q \pm i \widetilde{S}^2_q\right)/2$.

%%%%%%%%%%%%%%%%%%%%%%%%%%%%%%%%%%%%%%%%%%%
% commented out
\ifx
\begin{align}
    \widetilde{\chi}^{+-}(\qb,\omega) = \langle \widetilde{S}^+_{-\bs{q},-\omega}\widetilde{S}^-_{\bs{q},\omega}\rangle
\end{align}
or, in other terms, 
\begin{align}
    \widetilde{\chi}^{+-}(\bs{q},\omega) &= \frac{1}{4}\big[\widetilde{\chi}^{11}(\bs{q},\omega) -i\widetilde{\chi}^{12}(\bs{q},\omega)  \\ & +i \widetilde{\chi}^{21}(\bs{q},\omega) + \widetilde{\chi}^{22}(\bs{q},\omega)\big]
    \\
    \widetilde{\chi}^{-+}(\bs{q},\omega) &= \frac{1}{4}\big[\widetilde{\chi}^{11}(\bs{q},\omega) +i\widetilde{\chi}^{12}(\bs{q},\omega) \\ & -i \widetilde{\chi}^{21}(\bs{q},\omega) + \widetilde{\chi}^{22}(\bs{q},\omega)\big].
\end{align}
\fi
% commented out
%%%%%%%%%%%%%%%%%%%%%%%%%%%%%%%%%%%%%%%%%%%

%\PB{Commented out formulas that have already appeared in other paper }

The in- and out-of-plane temporal stiffnesses are then obtained from (see Eq.~\eqref{eq: chi dyn matrix form})
\begin{subequations}
\begin{align}
    Z^\smsqr &= \lim_{\omega \rightarrow 0}\chi^{33}(\bs{0},\bs{0},\omega), \\
    Z^\perp &= \lim_{\omega \rightarrow 0} 2\chi^{22}(\bs{0},\bs{0},\omega).
\end{align}
\label{eq: Z from chi}
\end{subequations}
Before discussing how to compute the susceptibility, we first write the Green's function in Eq.~\eqref{eq: Gf} within the rotated basis.  Defining an extended rotated basis as $\bs{\zeta}_{k} = \left( \psi_{k\up}, \psi_{k+Q,\down}, \bar{\psi}_{-k-Q,\up}, \bar{\psi}_{-k,\down}\right)^\mathrm{T}$, we get
%
%\begin{widetext}
%%%%%%%%%%%%%%%%%%%%%%%%%%%%
% commented out
\ifx
\begin{align}
    \widetilde{\mathcal{G}}^{-1}_{\bs{k},\nu} = 
    &\left(\begin{matrix}
        [i\nu - \xi_{\bs{k}}] & -\Deltam{k} & -\Delta_{t}^{\up}(\bs{k}) & \Deltap{k}  \\
        -\Deltam{k} & [i\nu - \xi_{\bs{k}+\bs{Q}}]  & -\Delta_{\rm p}(-\bs{k}-\bs{Q}) & -\Delta_{t}^{\down}(\bs{k}) \\
         -\Delta_{t}^{\up*}(\bs{k}) & -\Delta^*_{\rm p}(-\bs{k}-\bs{Q}) & [i\nu + \xi_{-\bs{k}-\bs{Q}}]  & \Delta_{{m}}(-\bs{k}-\bs{Q}) \\
        \Delta_{\rm p}^*(\bs{k}) & -\Delta_{t}^{\down *}(\bs{k}) & \Delta_{{m}}(-\bs{k}-\bs{Q}) & [i\nu + \xi_{-\bs{k}}]
    \end{matrix}\right)
\end{align}
\fi
% commented out
%%%%%%%%%%%%%%%%%%%%%%%%%%%%
\begin{equation}
\begin{split}
    &\widetilde{\mathcal{G}}^{-1}_{\bs{k}}(\nu_n) = \\
    &\left(\begin{matrix}
        i\nu_n - \xi_{\bs{k}} & -\Delta_{m} & 0 & -\Delta_{{p},\bs{k}}  \\
        -\Delta_{{m}} & i\nu_n - \xi_{\bs{k}+\bs{Q}} & \Delta_{{p},-\bs{k}-\bs{Q}} & 0 \\
         0 & \Delta_{{p},-\bs{k}-\bs{Q}} & i\nu_n + \xi_{-\bs{k}-\bs{Q}}  & \Delta_{{m}} \\
        -\Delta_{{p},\bs{k}} & 0 & \Delta_{{m}} & i\nu_n + \xi_{-\bs{k}}
    \end{matrix}\right).
\end{split}
\end{equation}

%$\equiv \left( \eta(k), \bar\eta(-k-Q) \right)$, where we defined $\eta(k) = \left( \psi_\up(k), \psi_\down(k+Q) \right)$. The $\eta(k)$ spinor, is the Fourier transform of the field in the rotated basis. 

In this basis, 16 different bilinears of the from $\bar{\zeta}_{k+q,a}\zeta_{k,b}$ ($a,b=1,...,4$) can be constructed, leading to $16\times16=256$ different susceptibilities of the form
\begin{equation}\label{eq: 256 full chis}
    \widetilde{\chi}^{abcd}(q) \sim \int_{k,k'} \left\langle \left(\bar{\zeta}_{k+q,a}\zeta_{k,b}\right) \left(\bar{\zeta}_{k',c}\zeta_{k'+q,d}\right)\right\rangle,
\end{equation}
To simplify the treatment and help physical intuition, we introduce a set of 16 linearly independent $4\times4$ matrices, $\Upsilon^\ell$, obeying
\begin{equation}
    \frac{1}{4}\mathrm{Tr}\left[\left(\Upsilon^\ell\right)^\dagger\Upsilon^{\ell'}\right] = \delta_{\ell,\ell'}.
\end{equation}
\\
These matrices can be arbitrarily chosen, and we define them such that (at least some of) the bilinears $(1/4) \bar{\zeta}_{k+q}\Upsilon^\ell \zeta_k$ acquire a physical meaning. We define the 16 Hermitian 4$\times$4 matrices $\Upsilon^\ell = \{\Gamma^{a}, A^a, \Phi^a, B^a \}$ with $l=0,\dots,15$ and
\begin{subequations}
\begin{align}
    \Gamma^{a} &= \left(\begin{matrix}
    \sigma^a & 0 \\
    0 & - \left(\sigma^a\right)^{\mathrm{T}} \end{matrix}\right), \\
     A^{a} &= \left(\begin{matrix}
    0 & t^a \\
    -t^a & 0 \end{matrix}\right),\\
    \Phi^{a} &= \left(\begin{matrix}
    0 & -i t^a \\
    -i t^a & 0 \end{matrix}\right),\\
    B^{a} &= \left(\begin{matrix}
    \sigma^a & 0 \\
    0 & \left(\sigma^a\right)^{\mathrm{T}} \end{matrix}\right),
\end{align}
\end{subequations}
\ifx
% OLD MATRICES
\begin{subequations}
\begin{align}
    \Gamma^{a} &= \left(\begin{matrix}
    \sigma^a & 0 \\
    0 & - \left(\sigma^a\right)^{\mathrm{T}} \end{matrix}\right), \\
     A^{a} &= \left(\begin{matrix}
    0 & t^a \\
    (t^a)^\dagger & 0 \end{matrix}\right),\\
    \Phi^{a} &= \left(\begin{matrix}
    0 & i t^a \\
    -i (t^a)^\dagger & 0 \end{matrix}\right),\\
    B^{a} &= \left(\begin{matrix}
    \sigma^a & 0 \\
    0 & \left(\sigma^a\right)^{\mathrm{T}} \end{matrix}\right),
\end{align}
\end{subequations}
\fi
%%%%%%%%%%%%%%% OLD EQUATION 
\ifx
\begin{subequations}
\begin{align}
    \Gamma^{a} &= \left(\begin{matrix}
    \sigma^a & 0 \\
    0 & - \left(\sigma^a\right)^{\mathrm{T}} \end{matrix}\right) \hskip3mm\text{if }\ell\in[0,3] \text{ $(a=\ell)$}, \\
     A^{a} &= \left(\begin{matrix}
    0 & t^a \\
    (t^a)^\dagger & 0 \end{matrix}\right) \hskip3mm\text{if }\ell\in[4,7]\text{ $(a=\ell-4)$},\\
    \Phi^{a} &= \left(\begin{matrix}
    0 & i t^a \\
    -i (t^a)^\dagger & 0 \end{matrix}\right) \hskip3mm\text{if }\ell\in[8,11] \text{ $(a=\ell-8)$},\\
    B^{a} &= \left(\begin{matrix}
    \sigma^a & 0 \\
    0 & \left(\sigma^a\right)^{\mathrm{T}} \end{matrix}\right) \hskip3mm\text{if }\ell\in[12,15] \text{ $(a=\ell-12)$},
\end{align}
\end{subequations}
\fi
%%%%%%%%%%%%%%%%%% END OLD EQUATION
%
with $a=0,1,2,3$ and $t^a = i\sigma^2\sigma^a$. 
With this choice, the $(1/4)\bar{\zeta}_{k+q}\Gamma^a\zeta_k$ takes the meaning of charge ($a=0$) and spin ($a=1,2,3$) fluctuations, while $(1/4)\bar{\zeta}_{k+q}A^a\zeta_k$ and $(1/4)\bar{\zeta}_{k+q}\Phi^a\zeta_k$ represent, in the case of singlet component ($a=0$), superconducting amplitude and phase fluctuations.  

%%%%%%%%%%%%%%%%%%%%%%%%%%%%%%%%%%%%%%%%%
% commented out
\ifx
, the spin bilinear can be re-expressed as
\begin{align}
    \tilde{S}^a_{\bs{q},\omega} & = \frac{1}{4} \int_{\bs{k}} \bar{\bs{\zeta}}_{k+q}\, \Gamma^a \,\bs{\zeta}_{k},
\end{align}
where we have defined the matrix
\begin{align}
    \Gamma^{a} = \left(\begin{matrix}
        \sigma^a & 0 \\
        0 & - \left(\sigma^a\right)^{\mathrm{T}} \end{matrix}\right).
\end{align}
To have a complete formula for the bare susceptibility and its vertex corrections, we have to find a complete set of bilinears for the coexisting magnetic and superconducting state. 
We use the complete set of bilinears
\begin{subequations}
\begin{align}
    \mathcal{S}^a_{k,q} &= \frac{1}{4} \bar{\bs{\zeta}}_{k+q}\, \Gamma^a\, \bs{\zeta}_{k},  \\
    \mathcal{A}^a_{k,q} &= \frac{1}{4} \bar{\bs{\zeta}}_{k+q} \,A^a\, \bs{\zeta}_{k},  \\
    \mathcal{P}^a_{k,q} &= \frac{1}{4} \bar{\bs{\zeta}}_{k+q}\, \Phi^a \,\bs{\zeta}_{k},  \\
    \mathcal{B}^a_{k,q} &= \frac{1}{4} \bar{\bs{\zeta}}_{k+q}\, B^a\, \bs{\zeta}_{k}.
\end{align}
\end{subequations}
where we have defined the 4x4 matrices ($\Gamma^a$ is as above)
\begin{subequations}
\begin{align}
    A^{a} &= \left(\begin{matrix}
        0 & t^a \\
        (t^a)^\dagger & 0 \end{matrix}\right), \\
    \Phi^{a} &= \left(\begin{matrix}
        0 & i t^a \\
        -i (t^a)^\dagger & 0 \end{matrix}\right), \\
            B^{a} &= \left(\begin{matrix}
        \sigma^a & 0 \\
        0 & \left(\sigma^a\right)^{\mathrm{T}} \end{matrix}\right) .
\end{align}
\end{subequations}
The bilinear $\mathcal{S}^a$ represent charge-spin fluctuations, $\mathcal{A}^0$ and $\mathcal{P}^0$ refers to amplitude and phase fluctuations for singlet pairing and $\mathcal{A}^i$ $\mathcal{P}^i$ for triplet pairing with $i=1,2,3$. It is important to underline here that the bilinear $\mathcal{B}^a$ has no $s$-wave component, i.e., $\int_{\bs{k}}\mathcal{B}^a(k)=0$. It is convenient to define the object $\Upsilon^l \equiv \{\Gamma^a, \Phi^0, A^0, \Phi^i, A^i, B^a\}$, which represents, in order, the charge, spin, pairing singlet, pairing triplet and $B^a$ vertices. 
% commented out
%%%%%%%%%%%%%%%%%%%%%%%%%%%%%%%%%%%%%%%%%
\fi
The susceptibilities~\eqref{eq: 256 full chis} can be conveniently re-expressed as
\begin{equation}
    \widetilde{\chi}^{\ell\ell'}(q) = \frac{1}{4}\sum_{abcd}\Upsilon^\ell_{ab}\,\widetilde{\chi}^{abcd}(q)\Upsilon^\ell_{cd}.
\end{equation}

\ifx
%%%%%%%%%%%%%%%%%%%%%%%%%%%%%%%%%%%%%%%%%
% commented out
By using such complete basis, one can construct all possible correlators, that are possibly non-zero in the coexisting phase. Indeed, one can define the vertex function as
\begin{align}
    \widetilde{V}^{ll'}(k_1,k_2,k_3) = \langle \bar{\bs{\zeta}}(k_1)\Upsilon^l\bs{\zeta}(k_2) \bar{\bs{\zeta}}(k_3)\Upsilon^{l'}\bs{\zeta}(k_1+k_2-k_3) \rangle_{\rm 1PI}
\end{align}
Such vertex function contains already normal and anomalous components in both magnetic and superconducting channels, all terms that violates SU(2) and U(1) symmetries; hence, also terms like $\psi\psi\psi\bar{\psi}$, for instance. 
\DV{I want to comment that the resulting vertex is the most general one in the coexisting phase. Maybe also write an introductory formula for it. }

To compute the vertex corrections, in principle all susceptibilities constructed from these bilinears are important. In the coexisting phase, the general susceptibility according to the actions~\eqref{eq: S} and~\eqref{eq: expA}, in the rotated basis, takes the form
\begin{align}
\label{eq: tilde chi}
    \widetilde{\chi}^{ll'}(\qb,\omega) = \chi^{ll'}_0(\qb,\omega) + \sum_{mn} \widetilde{\chi}^{ln}_0(\qb,\omega) \widetilde{V}^{nm}(\qb,\omega) \widetilde{\chi}^{ml'}(\qb,\omega)
\end{align}
with $l=1,\dots,16$. Here, $\widetilde{V}^{nm}(\qb,\omega)$ is the irreducible vertex in the rotated basis. 
%  commented out
%%%%%%%%%%%%%%%%%%%%%%%%%%%%%%%%%%%%%%%%%
\fi

Within the RPA, $\widetilde{\chi}(q)$ as a matrix in $\ell$, $\ell'$ is computed as %
\begin{equation}\label{eq: chi RPA}
\begin{split}
    \widetilde{\chi}(q) =  \widetilde{\chi}_0(q)\left[\mathbb{1}-\widetilde{V}(\bs{q})\widetilde{\chi}_0(q)\right]^{-1}.
\end{split}
\end{equation}
In the above equation, we have assumed the effective interaction $U^{\ell\ell'}_{\bs{k}\bs{k}'}(\bs{q})$ to be factorizable, namely that
\begin{equation}
    U^{\ell\ell'}_{\bs{k}\bs{k}'}(\bs{q}) = \widetilde{V}^{\ell\ell'}(\bs{q}) f^\ell_\bs{k} f^{\ell'}_{\bs{k}'},
\end{equation}
with $f^\ell_\bs{k}$ some model-dependent form factors. For the specific case under study, we have
\begin{equation}
\label{eq: form factors}
    f^\ell_\bs{k} = 
    \begin{cases}
        &1 \hskip 5mm\text{if } \ell\in[0,3] \\    
        &\cos k_x-\cos k_y \hskip 5mm\text{if } \ell=4\text{ or }\ell=8 \\    
        & 0 \hskip 5mm\text{otherwise}, \\    
    \end{cases}
\end{equation}
while the functions $\widetilde{V}^{\ell\ell'}(\bs{q})$ read as
%\begin{widetext}
%\begin{align}
%\label{eq: Interaction}
%    \widetilde{V}(\qb) = \left(\begin{matrix}
%        -U_c(\qb) & 0 & 0 & 0 & 0 & 0 \\
%         0 & U_m(\qb-\QQ) & 0 & 0 & 0 & 0 \\
%         0  & 0 & U_m(\qb-\QQ) & 0 & 0 & 0 \\
%         0 & 0 & 0 & U_m(\qb) & 0 & 0 \\
%         0 & 0 & 0 & 0 & U_p(\qb) & 0 \\
%         0 & 0 & 0 & 0 & 0 & U_p(\qb) 
%    \end{matrix}
%    \right)
%\end{align}
\begin{equation}\label{eq: Interaction}
    \begin{split}
        \widetilde{V}(\qb) = \mathrm{diag}\Big(
       -U_c(\qb),\, &U_m(\qb-\QQ),\, U_m(\qb-\QQ) \\
       &U_m(\qb),\,  U_p(\qb),\, U_p(\qb)\Big)\,,
    \end{split}
\end{equation}
%\end{widetext} 
for $l,l'=0,\dots,6$, while all the other components are zero. 
\ifx
\begin{equation}
    \widetilde{V}^{\ell\ell'}(\bs{q}) = 
    \begin{cases}
        &-U_\mathrm{c}(\bs{q})\hskip5mm\text{if }\ell=\ell'=0\\
        &U_{m}(\bs{q})\hskip5mm\text{if }\ell=\ell'=1,2,3\\
        &U_{p}(\bs{q})\hskip5mm\text{if }\ell=\ell'=4\text{ or }8\\
        &0\hskip5mm\text{otherwise}.
    \end{cases}
\end{equation}
\fi
The bare susceptibilities or bubbles $\widetilde{\chi}_0^{\ell\ell'}(q)$ appearing in Eq.~\eqref{eq: chi RPA} are given by
\begin{equation}
\label{eq: rotated bubble}
\begin{split}
    \widetilde{\chi}^{\ell\ell'}_0(\bs{q},\omega) &= -\frac{1}{4}\int_{\bs{k}} f^\ell_\bs{k}f^{\ell'}_\bs{k} T\sum_{\nu_n}\Tr 
    \Big[ 
    \Upsilon^\ell \widetilde{\mathcal{G}}_{\bs{k}}(\nu_n) \\
    &\times\Upsilon^{\ell'} \widetilde{\mathcal{G}}_{\bs{k}+\bs{q}}(\nu_n+\Omega_m) \Big]\Big|_{i\Omega_m \rightarrow \omega +i0^+}.
\end{split}
\end{equation}
The Matsubara summation above can be carried out analytically as described in Appendix~\ref{app: Matsubara}. 

To summarize, the temporal stiffnesses, as in Eq.~\eqref{eq: Z from chi}, are calculated by combining Eqs.~\eqref{eq: chi RPA} and \eqref{eq: chi22 orig to rot} for the susceptibility. 

\subsection{Gap equations}

The gaps are computed via a renormalized set of mean-field gap equations~\cite{Eberlein2016,Vilardi2020} 
\begin{align}
\label{eq: gap m}
    \Delta_{{m}} &= - U_{{m}}(\QQ)\int_{\kk} T\sum_\nu F^{{m}}_{\kk}(\nu) \\
\label{eq: gap sc}
    \Delta_{\rm p} &=  U_{\rm p}(\bs{0}) \int_{\kk} T\sum_\nu d_{\kk} F^{\rm p}_{\kk}(\nu)
\end{align}
where the full superconducting gap $\Delta_{\rm p}(\kk) = d_\kk \Delta_{\rm p}$, Fourier transform of  $\Delta_{p, jj'}$, has been decomposed into d-wave form factor $d_\kk = \cos{k_x} - \cos{k_y}$ as a consequence of the decomposition~\eqref{eq: form factors}. The interactions $U_m$ and $U_{\rm p}$ will be computed via some functional renormalization group method.  

Here, we simplify the momentum dependence of the gap function via a simple form-factor projection: $s$-wave sector for the magnetic gap and $d$-wave for the superconducting gap.  

The wave vector $\QQ$ is determined by minimizing the Free energy, which is given by
\begin{align}
    F(\QQ) =  -\frac{1}{2} T \int_\kk \sum_l \ln{} &\left( 1 + e^{-E^l_\kk /T} \right) + \frac{1}{2}\int_\kk \left(\epsilon_\kk + \epsilon_{\kk + \QQ}\right) \nonumber\\
    & + \frac{\Delta_m^2}{U_m(\QQ)} + \frac{\Delta_{\rm p}^2} {U_p(\mathbf{0})} + \mu n \ .
\end{align}
$E^l_\kk$ are the four eigenvalues of the Hamiltonian~\eqref{eq: Ham}. An exact analytical formula for $E^l_\kk$ can be found in Ref.~\cite{Metzner2019}. 

\section{Numerical results}
\label{sec: results}

\begin{figure}[t]
    \centering
    \includegraphics[width=0.46\textwidth]{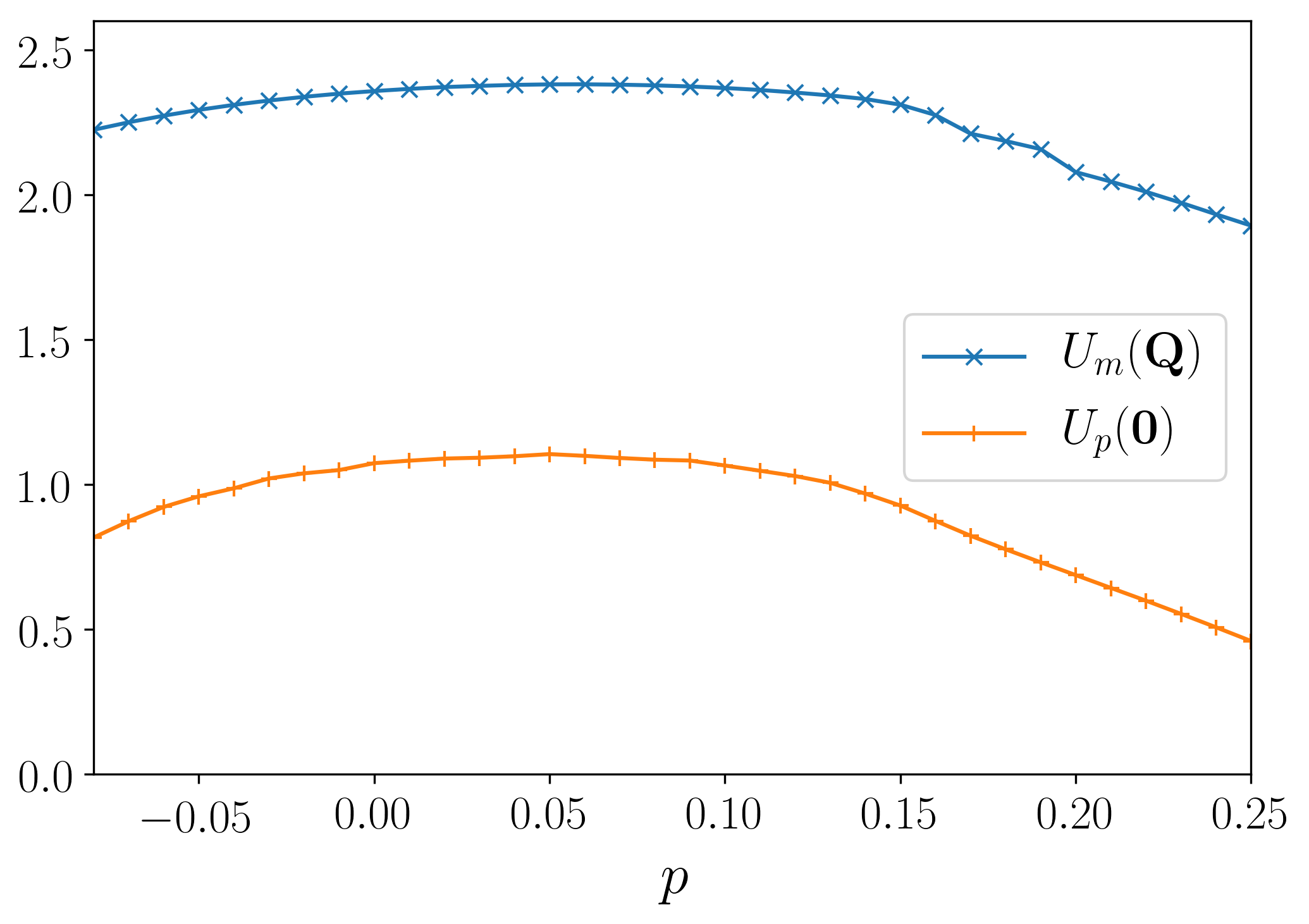}
    \caption{Magnetic and pairing effective interactions as a function of the doping. }
\label{fig: Eff Int}
\end{figure}

\begin{figure}[t]
    \centering
    \includegraphics[width=0.46\textwidth]{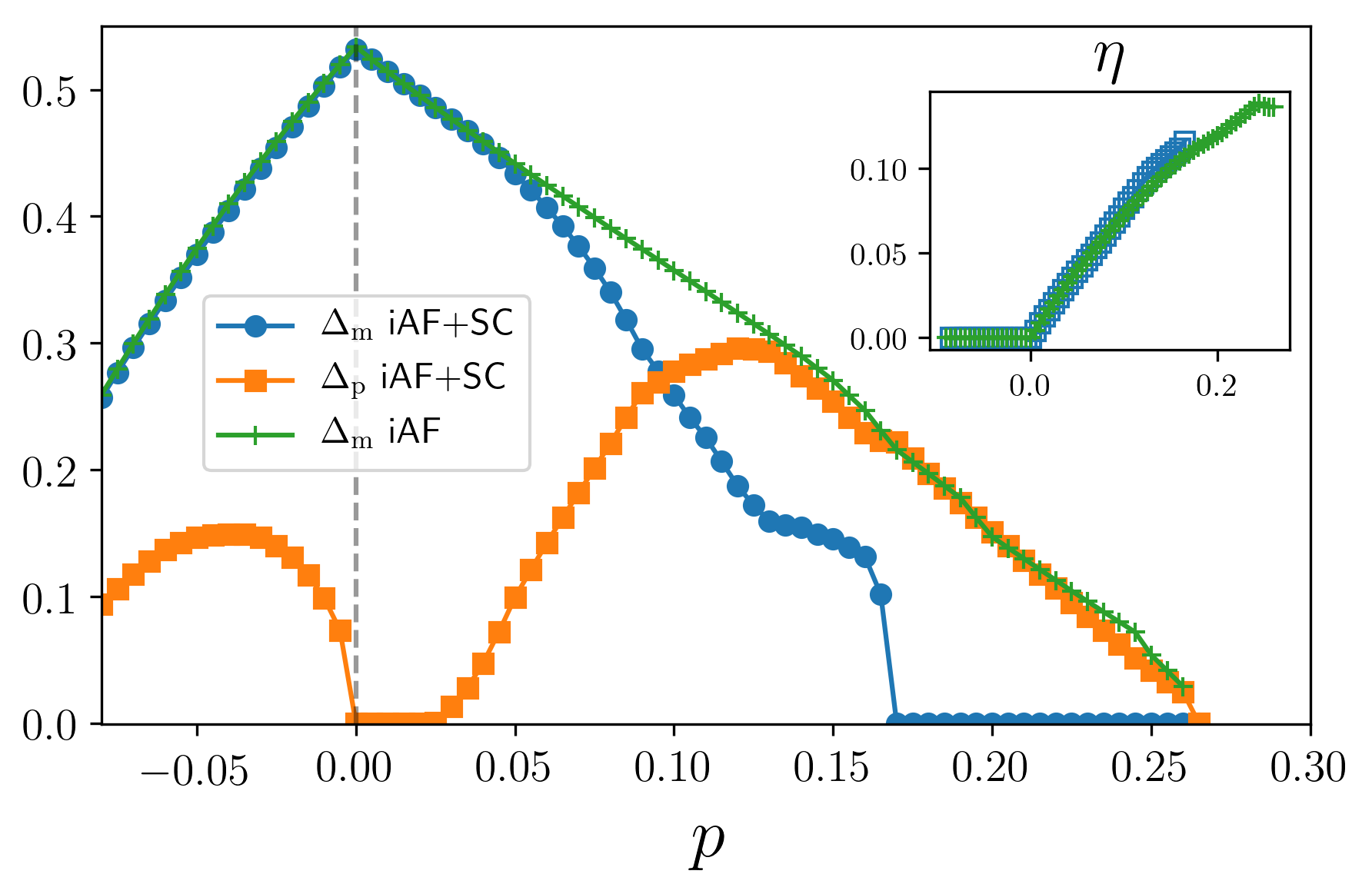}
    \caption{Superconducting and magnetic gaps as functions of the doping $p=1-n$ at temperature $T=0.001t$. $\Delta_{\rm m}$ and $\Delta_{\rm p}$ in blue and orange, respectively, refer to the solution with coexisting magnetism and superconductivity. Conversely, the green data refers to the magnetic solution without the superconducting phase. Inset: incommensurability $\eta$ as a function of the doping for the magnetic solution (in green) and for the superconducting magnet one (in blue). }
    \label{fig: gaps}
\end{figure}

\begin{figure}[t]
    \centering
    \includegraphics[width=0.46\textwidth]{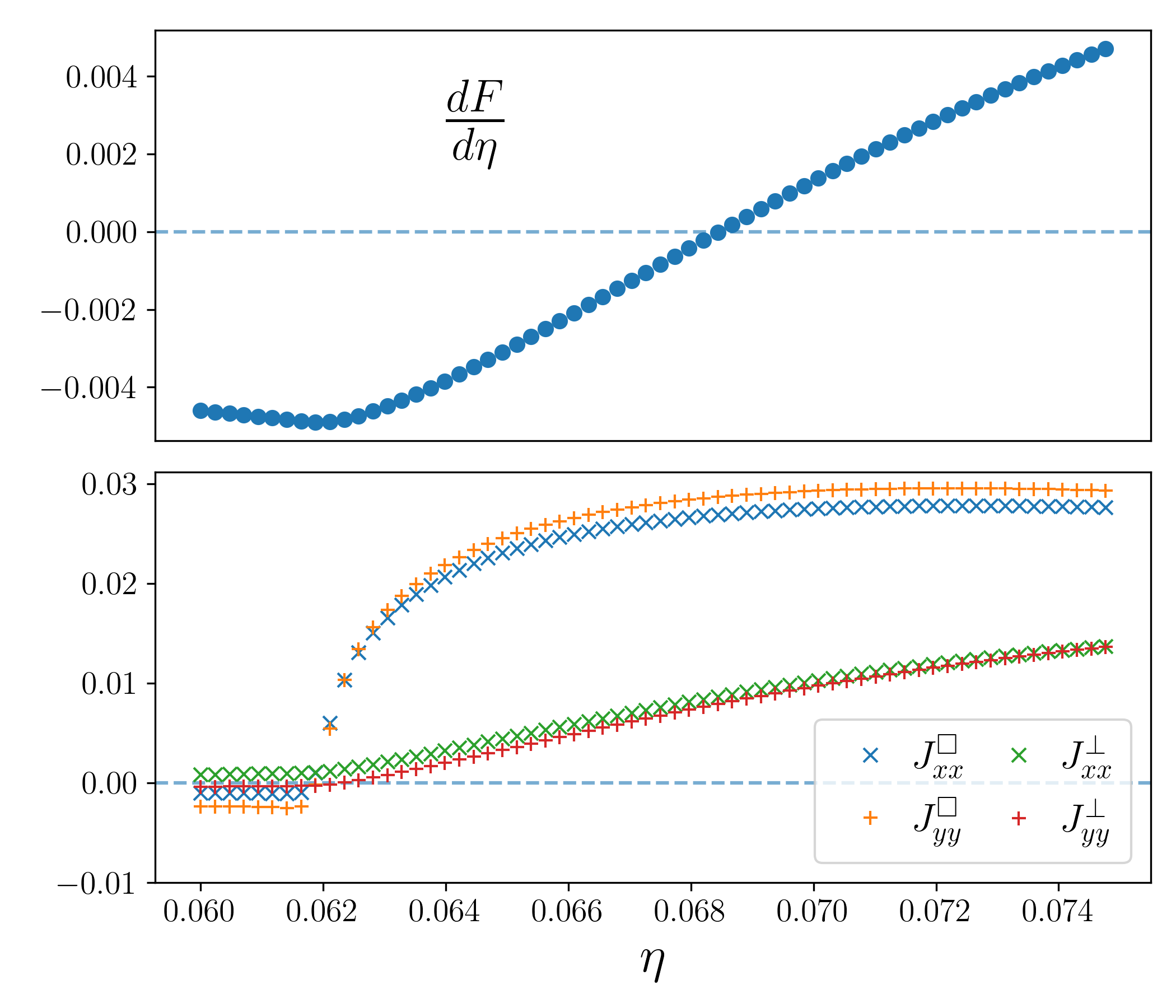}
    \caption{Upper panel: derivative of the free energy shown as a function of the incommensurability factor $\eta$ for $n=0.91$. Lower panel: out-of-plane $J^{\perp}$ and in-plane $J^{\smsqr}$ stiffnesses as a function of $\eta$. }
\label{fig: dFdeta}
\end{figure}

In this section, we show the spatial and temporal stiffnesses computed with nearest and next-to-nearest hopping amplitudes respectively $t$ and $t'=-0.2t$. We fix the energy units by imposing $t=1$. The gap functions defined in the Hamiltonian~\eqref{eq: Ham} are computed through Eqs.~\eqref{eq: gap m} and~\eqref{eq: gap sc}. The effective couplings are calculated via functional renormization group~\cite{Eberlein2016,Bonetti2022Gauge}, where the high energy modes are integrated out and effective low energy couplings can be obtained in all interacting channels. As in Ref.~\cite{Bonetti2022Gauge}, we use a temperature flow~\cite{Honerkamp2001} starting with a bare coupling $U=4t$. More details can be found in~\cite{Eberlein2016,Bonetti2022Gauge}. In Fig.~\ref{fig: Eff Int} we show both the magnetic and pairing effective interactions as a function of the doping $p=1-n$. 

\subsubsection{Gaps}

In Fig.~\ref{fig: gaps} we show the magnetic and superconducting gaps as a function of the doping computed at finite but small temperature $T=0.001t$. We confirm the already found coexisting order for a wide doping regime, as also found in Ref.~\cite{Eberlein2016,Vilardi2020,Metzner2019}. For comparison, we also plot the antiferromagnetic solution (without superconductivity), which is always found to be higher in energy.

The magnetic gap peaks at half-filling and vanishes around $n=0.83$ at the van Hove singularity. Consistent with Ref.~\cite{Eberlein2016}, a coexisting solution was also observed for $n<0.83$; however, we exclusively show here only the superconducting state. This is because the energy difference between the coexisting and purely superconducting states is negligible ($10^{-6}t$), indicating a fragile magnetic order and that the energy gain is primarily due to superconductivity. 
%\DV{(Once Robin's paper is on arxiv I can cite it and mention that for $n<0.83$ we expect stripe order and that the formula for the spin stiffness is different.)}

%The magnetic gap is largest at half-filling and vanishes around $n=0.83$ at the van Hove singularity, consistent with Ref.~\cite{Eberlein2016}. For $n<0.83$, we present only the superconducting solution because the energy difference between the coexisting and purely superconducting states is negligible ($10^{-6}t$), indicating a fragile magnetic order and that the energy gain is primarily due to superconductivity.

\begin{figure}[t]
    \centering
    \includegraphics[width=\linewidth]{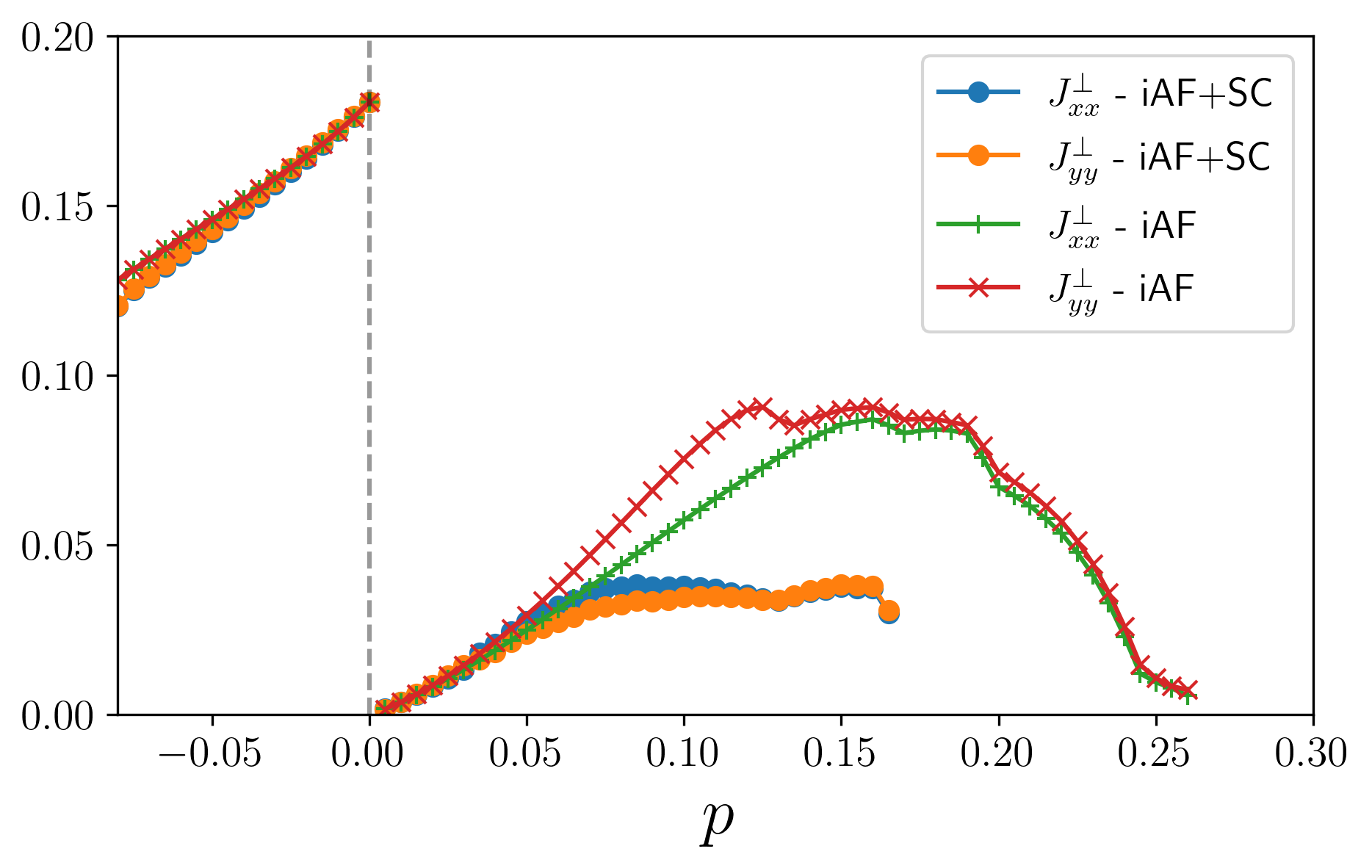}
    \caption{Out-of-plane stiffnesses as a function of the doping $p=1-n$. In blue and orange the stiffnesses for the coexistence and in red and green for the purely magnetic state. }
    \label{fig: Jo}
\end{figure}
\begin{figure}[t]
    \centering
    \includegraphics[width=\linewidth]{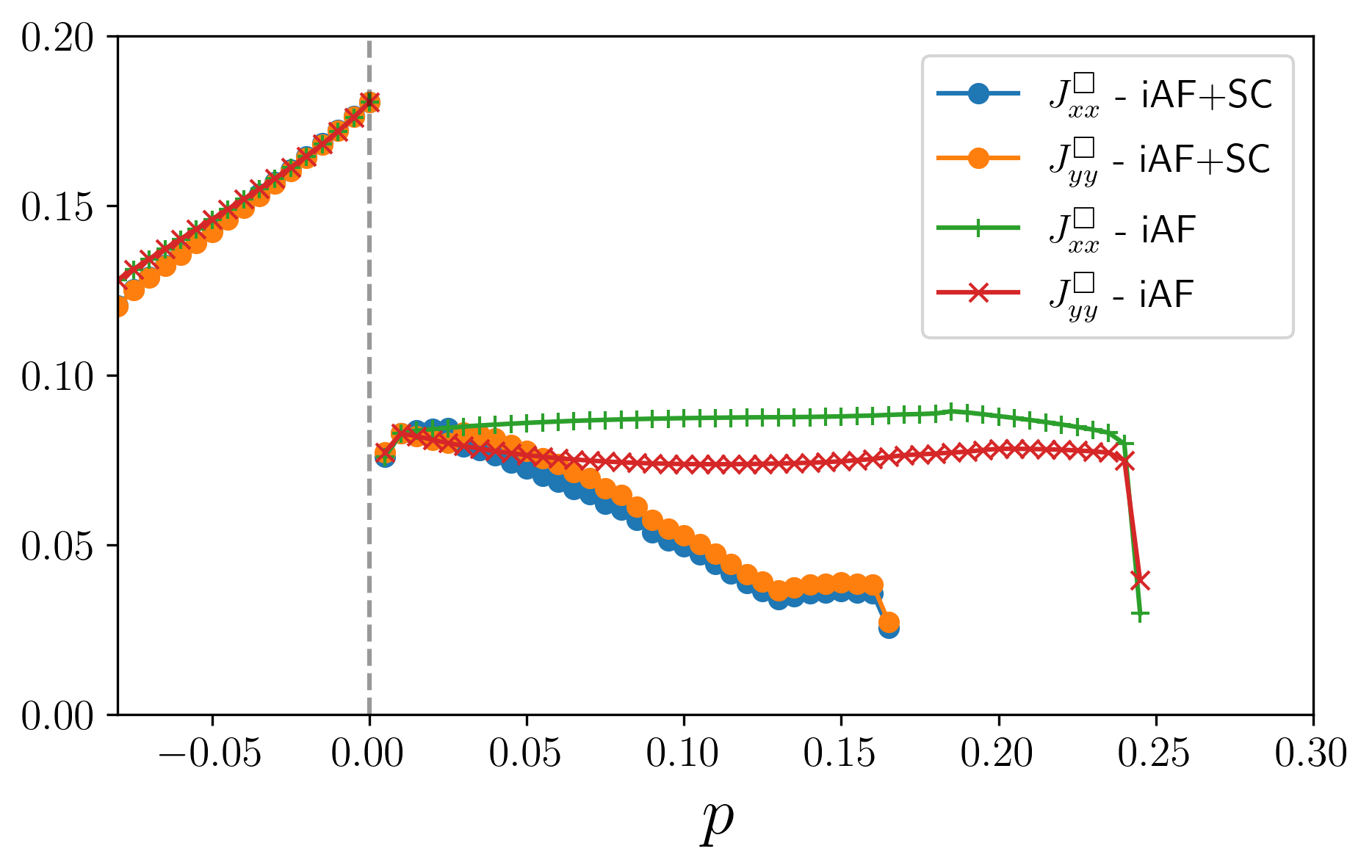}
    \caption{In-plane stiffness as a function of the doping $p=1-n$. In blue and orange the stiffnesses for the coexistence and in red and green for the purely magnetic state.}
    \label{fig: Ji}
\end{figure}
\begin{figure}[t]
 \centering
 \includegraphics[width=\linewidth]{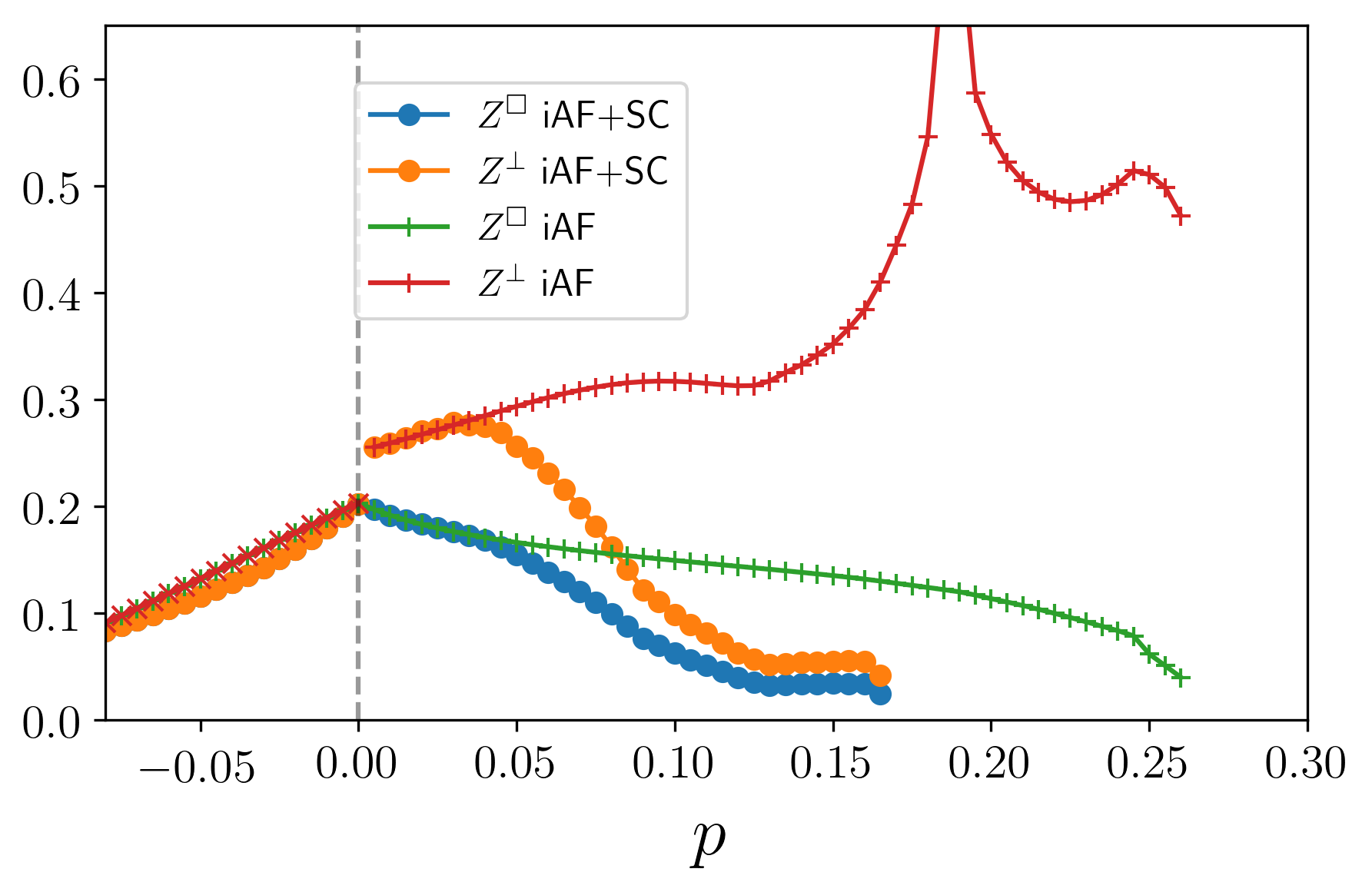}
 \caption{In-plane and out-of-plane temporal stiffnesses as a function of doping. }
\label{fig: Z}
\end{figure}

The superconducting gap behavior at small doping is different for the hole-doped and the electron-doped cases, as also found in~\cite{Metzner2019}.  For hole-doping, once the superconducting gap becomes sizable, it reduces the magnetization. In contrast, the superconducting gap emerges rapidly upon the introduction of electron doping, while the magnetic gap remains practically unaffected by the presence of the superconducting order itself.

Regarding the specific type of magnetic ordering, we found the \Neel \ state in the electron-doped case, while for hole doping the magnetic order is an incommensurate spiral with momentum vector $\QQ=(\pi,\pi-2\pi\eta)$. The incommensurability $\eta$, shown in the inset of Fig.~\ref{fig: gaps}, is finite for any doping value $p>0$. 

The correct spiral momentum vector $\QQ$ is determined via the minimization of the free energy. In practice, we find the zero of the function $\frac{dF}{d\eta}=-2\pi\frac{dF}{dQ_y}$, which is shown in Fig.~\ref{fig: dFdeta} as a function of $\eta$. It is worthwhile to remark that the second derivative $\frac{d^2 F}{dQ^2_y}$ equals the in-plane stiffness $J^{\smsqr}_{yy}$~\cite{Bonetti2022Ward}, which is also shown in the lower plot. In fact, when the derivative $\frac{dF}{d\eta}$, in the upper plot, changes slope, at around $\eta\approx 0.62$, the in-plane stiffness drops to very small values. In this scenario, the N\'eel state ($\eta=0$) constitutes a maximum of the free energy, or $\frac{dF}{d\eta}(\eta=0)=0$, hence, the derivative $\frac{dF}{d\eta}$ needs to be negative as $\eta \rightarrow 0$. 

\subsubsection{Stiffnesses}

Figure~\ref{fig: Jo} presents a comprehensive scan of the out-of-plane spin stiffness as a function of the electron and hole doping. At small hole-doping, where the superconducting gap is still small, the stiffness of the superconducting state closely matches that of the pure magnetic state, indicating that superconductivity has a minimal initial impact on the magnetic correlations. However, for larger dopings, as the superconducting order strengthens, the spin stiffness of the superconducting state proves significantly smaller than that of the pure magnetic state. This suppression continues as doping increases, a trend that persists until the magnetic order completely stops at the van Hove filling ($n=0.83$).

This pronounced reduction in stiffness at larger hole-doping is attributed to an interplay of three principal effects. First, and most significantly, the emergence of the superconducting state directly suppresses the underlying magnetic gap. This suppression, in turn, leads to a smaller contribution from the terms in Eqs.~\eqref{eq: J out} and \eqref{eq: J in} which are proportional to the normal Green's function $G_\kk(\nu)$ and the magnetic anomalous Green's function $F^m_\kk(\nu)$. These terms, representing contributions also present in a purely magnetic state, are diminished not only due to the reduced magnetic gap but also because the finite superconducting gap further suppresses the values of both $G_\kk(\nu)$ and $F^m_\kk(\nu)$. Secondly, novel terms arise that are proportional to the superconducting anomalous Green's functions $F^s_\kk(\nu)$ and $F^t_\kk(\nu)$. These additional terms directly contribute to and further reduce the value of the final stiffness. Thirdly, while potentially contributing to stiffness suppression, superconducting vertex corrections, defined as $\Delta^\alpha_\kk=\partial_\kk \Delta_\kk$, were numerically checked and confirmed to be marginal (less than $10^{-5}$), suggesting their impact is negligible. Interestingly, on the electron-doped side, this difference in stiffness between the superconducting and pure magnetic states becomes extremely small, almost negligible, despite the rapid emergence of the superconducting gap with electron doping. 

Figure~\ref{fig: Ji} presents the in-plane stiffness, which qualitatively replicates the behavior of the out-of-plane stiffness. It exhibits a similar suppression by the superconducting gap for hole-doping, confirming that the impact of superconductivity on magnetic correlations is generally strong in this regime. Conversely, also the in-plane stiffness remains largely unaffected for electron-doping.

The temporal stiffness, shown in Fig.~\ref{fig: Z}, is a crucial indicator of the dynamics of spin fluctuations and the overall stability of magnetic order. At approximately $p\simeq 19\%$, the purely magnetic solution exhibits a prominent divergence, which is attributed to a van Hove singularity appearing in the quasi-particle band structure~\cite{Bonetti2022Gauge}. In stark contrast to this purely magnetic scenario, the coexisting solution, where superconductivity is present, terminates at the van Hove singularity of the bare dispersion, located at a different doping value of $p=17\%$. This distinction in the termination points underscores how the interplay with superconductivity significantly modifies the underlying electronic structure and its influence on magnetic stability, potentially leading to a quantum disordered state in the absence of long-range magnetic order.

%%%%%%%%%%%%%%%%%%%%%%%%%%%%%%%%%%%%%%%%%%%%%%%%%%
%
%      CONCLUSIONI
%%%%%%%%%%%%%%%%%%%%%%%%%%%%%%%%%%%%%%%%%%%%%%%%%%
\section{Conclusions}
\label{sec: conclusions}

In this study, within the framework of an SU(2) gauge theory for the pseudogap phase, we use the fractionalization of the electron field into spinons and chargons~\cite{Scheurer2018} to include superconductivity. Specifically, we considered the superconducting instability, named the SC$^*$ phase, arising from the pairing between chargons. Our analysis reveals that the saddle-point solution for the chargon naturally accommodates a coexisting phase with both magnetic and superconducting orders, a finding consistent with prior observations from functional renormalization group improved gap equations~\cite{Eberlein2016,Vilardi2020}. Our primary objective was to address how the superconducting state influences the strength of quantum fluctuations, a property quantitatively assessed via the spin stiffnesses. This work is complementary to previous studies that have focused on the effect of magnetic order on the superconducting phase stiffness~\cite{Metzner2019}.

We have rigorously derived both the spatial and temporal spin stiffnesses for a state characterized by coexisting magnetic and superconducting order. These were obtained as response functions to an external SU(2) gauge field. A crucial aspect of our theoretical formulation is the explicit inclusion of the coupling between the gauge field and the non-local superconducting gap~\cite{Verma2021}. This coupling introduces, in addition to the standard current vertex, a term proportional to the gradient of the gap function, which proved essential for ensuring the restoration of gauge symmetry within our formalism.

Our calculations, for the specific coupling values employed, demonstrate that the system exhibits an incommensurate spiral order at any finite hole-doping, while preserving N\'eel order at electron-dopings and at half-filling. We have reproduced the previously reported jump in the stiffnesses at low hole-doping, as noted in Ref~\cite{Bonetti2022Gauge,Vilardi2025a}. A significant outcome of this work is the finding that at large hole-doping, the presence of the superconducting gap actively enhances spin fluctuations by suppressing the spin stiffnesses. This reduction in spin stiffnesses suggests that superconductivity can play a role in destabilizing long-range magnetic order, thereby promoting and potentially stabilizing the topological order.

Our framework thus provides a valuable theoretical laboratory for exploring the complex interplay of competing orders in systems that might host fractionalized phases. The insights gained into the reduction of spin stiffness by superconductivity contribute to the broader understanding of quantum critical phenomena and the conditions under which topological order can persist. Furthermore, a useful byproduct of our calculation is that the derived formulas for the spin stiffnesses are generally applicable to more conventional systems where magnetic and superconducting orders coexist, allowing for a broader application of our results beyond the specific fractionalized context. 

\section*{Acknowledgements}
We are thankful to Walter Metzner, Andrey Chubukov and Subir Sachdev, for enlightening discussions. We are also grateful to Walter Metzner for carefully reading our manuscript. 
P.M.B. acknowledges support by the German National Academy of Sciences Leopoldina through Grant No. LPDS 2023-06.

%%%%%%%%%%%%%%%%%%%%%%%%%%%%%%%%%%%%%%%%%%%%%%%%%%
%%%%%%%%%%%%%%%%%%%%%%%%%%%%%%%%%%%%%%%%%%%%%%%%%%
%%%%%%%%%%%%%%%%%%%%%%%%%%%%%%%%%%%%%%%%%%%%%%%%%%
\begin{appendix}
\begin{widetext}
\section{Expressions for the paramagnetic and diamagnetic terms}
\label{app: param and diam terms}
Here we report the complete formulas for the in-plane and out-of-plance stiffnesses. 
\subsubsection{In-plane mode}
For the in-plane spin stiffness, we find the following contributions
\begin{subequations}
\begin{align}
    & J^{\smsqr,\Gamma\Gamma}_{\alpha\beta} = \frac{1}{2}
    \int_{\bs{k}} T\sum_{\nu_n} \Big\{
    \gamma^\alpha_{\bs{k}}\gamma^\beta_{\bs{k}} G^2_{\bs{k}}(\nu_n) 
    - \gamma^\alpha_{\bs{k}}\gamma^\beta_{\bs{k}} [F^{s}_{\bs{k}}(\nu_n)]^2 
    -\gamma^\alpha_{\bs{k}}\gamma^\beta_{\bs{k}+\bs{Q}} [F^{{m}}_{\bs{k}}(\nu_n)]^2 
    +\gamma^\alpha_{\bs{k}+\bs{Q}} \gamma^\beta_{\bs{k}} [F^{t}_{\bs{k}}(\nu_n)]^2
    \Big\},\\
%%%%%%%%%%%%%%%%%%%%%%%%%%%%%%%%%%%
    & J^{\smsqr,DD}_{\alpha\beta} = 
    \frac{1}{2} \int_{\bs{k}} T\sum_{\nu_n} 
    \Big\{ 
    -\Delta^\alpha_{\bs{k}}\Delta^\beta_{\bs{k}} |G_\bs{k}(\nu_n)|^2
    +\Delta^\alpha_{\bs{k}}\Delta^\beta_{\bs{k}} [F^s_\bs{k}(\nu_n)]^2
    -\Delta^\alpha_{\bs{k}}\Delta^\beta_{\bs{k}+\bs{Q}} |F^{{{m}}}_{\bs{k}}(\nu_n)|^2
    +\Delta^\alpha_{\bs{k}}\Delta^\beta_{\bs{k+Q}}
    |F^t_\bs{k}(\nu_n)|^2
    \Big\},\\
%%%%%%%%%%%%%%%%%%%%%%%%%%%%%%%%%%%%%
    & J^{\smsqr,\Gamma D}_{\alpha\beta} = 
    \int_{\bs{k}}T\sum_{\nu_n} \Big\{
    \gamma^\alpha_{\bs{k}}\Delta^{\beta}_{\bs{k}} G_\bs{k}(\nu_n)F^s_\bs{k}(\nu_n) 
    +\gamma^\alpha_{\bs{k}}\Delta^{\beta}_{\bs{k}+\bs{Q}} F^m_\bs{k}(\nu_n)F^t_\bs{k}(\nu_n) 
    \Big\}.   
\end{align}
\end{subequations}
The diamagnetic term reads as
\begin{equation}\label{eq: J diag}
\begin{split}
    J^{\rm d}_{\alpha\beta} &= \frac{1}{2}\int_{\kk} T\sum_{\nu_n}\left\{ \gamma^{\alpha\beta}_{\bs{k}} G_{\bs{k}}(\nu_n) 
    +\Delta^{\alpha\beta}_{\bs{k}}  F^s_{\bs{k}}(\nu_n)\right\}.
\end{split}
\end{equation}
\subsubsection{Out-of-plane stiffness}
For the out-of-plane stiffness, we find instead
\begin{subequations}
\begin{align}
    & J^{\perp,\Gamma\Gamma}_{\alpha\beta} = \frac{1}{2}
    \int_{\bs{k}} T\sum_{\nu_n} \Big\{
    \gamma^\alpha_{\bs{k}}\gamma^\beta_{\bs{k}}
    G_{\bs{k}}(\nu_n)G_{-\bs{k}}(\nu_n) 
    -\gamma^\alpha_{\bs{k}}\gamma^\beta_{\bs{k}}
    F^s_{\bs{k}}(\nu_n)F^s_{-\bs{k}}(\nu_n) 
    \Big\}, \\
%%%%%%%%%%%%%%%%%%%%%%%%%%%%%%%%%%%%%%
    &J^{\perp,DD}_{\alpha\beta} = \frac{1}{2}
    \int_{\bs{k}} T\sum_{\nu_n}
    \Big\{
    -\Delta^\alpha_{\bs{k}}\Delta^\beta_{\bs{k}}
    G_{\bs{k}}(\nu_n)G^*_{-\bs{k}}(\nu_n) 
    +\Delta^\alpha_{\bs{k}}\Delta^\beta_{\bs{k}}
    F^s_{\bs{k}}(\nu_n)F^s_{-\bs{k}}(\nu_n) 
    \Big\},\\
%%%%%%%%%%%%%%%%%%%%%%%%%%%%%%%%%%%%%%
    & J^{\perp,\Gamma D}_{\alpha\beta} = 
    \int_{\bs{k}}T\sum_{\nu_n} \Big\{
    \gamma^\alpha_{\bs{k}}\Delta^{\beta}_{\bs{k}}G_{-\bs{k}}(\nu_n)F^s_\bs{k}(\nu_n)
    \Big\}.
\end{align}
\end{subequations}
\end{widetext}
%
%%%%

\section{Matsubara summation}
\label{app: Matsubara}

Here, we show how we perform the Matsubara summation for the stiffnesses in Eqs.~\eqref{eq: J out} and~\eqref{eq: J in}, as well as for the bubble~\eqref{eq: rotated bubble}. We start considering the latter case, which involves the Green's function in the rotated basis. Due to its translational invariance, it can be simply written as 
\begin{align}
    \mathcal{\widetilde{G}}_{\bs{k}}(\nu) = U^\dagger_{\bs{k}}{\rm diag}\left(\frac{1}{i\nu - E^l_{\bs{k}}}\right) U_{\bs{k}}
\label{eq: G dec}
\end{align}
where $E^l_{\bs{k}}$ are the four eigenvalues of the Hamiltonian, which, in this context, is $\hat H_\kk = i\nu \mathbb{1} - \mathcal{\widetilde{G}}_\kk$. $U_{\bs{k}}$ is the unitary transformation matrix that diagonalize the Green's function. By using~\eqref{eq: G dec}, the Matsubara summation in the bubble~\eqref{eq: rotated bubble} can be carried giving rise to
\begin{align}
\begin{split}
    \widetilde{\chi}^{ab}_0(\bs{q},\omega) = -\frac{1}{16}\int_{\bs{k}}
    &\sum_{l,l'} \Tr \left[ \Gamma^a \bs{u}_{l,\bs{k}} \Gamma^b \bs{u}_{l',\bs{k}+\bs{q}} \right] \\ \times &\frac{f(E^l_{\bs{k}}) - f(E^{l'}_{\bs{k}+\bs{q}})}{\omega + i0^+ + E^l_{\bs{k}} - E^{l'}_{\bs{k}+\bs{q}}}
\end{split}
\end{align}
where we defined $u^{\alpha\beta}_{l,\bs{k}} = U^{*\alpha}_{l,\bs{k}} U_{l,\bs{k}}^\beta$ and $U_{l,\bs{k}}^\beta = \left(U_{\bs{k}}\right)_{l\beta}$ are the matrix element of the transformation matrix $U_\bs{k}$. 

\ifx
The Green's function can be decomposed as 
\begin{align}
\label{eq: tGdecomp}
    \mathcal{\widetilde{G}}^{\alpha\beta}_{\bs{k}}(\nu) = \sum_l \frac{u^{\alpha\beta}_{l,\bs{k}}}{i\nu - E^l_{\bs{k}}} 
\end{align}
where we defined $u^{\alpha\beta}_{l,\bs{k}} = U^{*\alpha}_{l,\bs{k}} U_{l,\bs{k}}^\beta$  and $U_{l,\bs{k}}^\beta = \left(U_{\bs{k}}\right)_{l\beta}$ are the matrix elements of the transformation matrix, and are called strutturini di formina.
\fi 

Regarding the Matsubara summation in the calculation of the stiffness, Eqs.~\eqref{eq: J in} and~\eqref{eq: J out}, We first note that the relations between the rotated and unrotated Green's function components, for instance, 
\begin{align*}
    F^{\rm m}_{\bs{k}}(\nu) &= \widetilde{F}^{\rm m}_{\bs{k}}(\nu) & 
    \bar{F}^{\rm m}_{\bs{k}}(\nu) &= \widetilde{\bar{F}}^{\rm m}_{\bs{k}}(\nu) \\
    G_{\bs{k}}(\nu) &= \widetilde{G}_{\bs{k}}(\nu) &
    \bar{G}_{\bs{k}-\QQ}(\nu) &= \widetilde{\bar{G}}_{\bs{k}-\bs{Q}}(\nu) \\
    F^{t \down}_{\bs{k}}(\nu) &= \widetilde{F}^{t \down}_{\bs{k}-\bs{Q}}(\nu) &   
    F^{t \up}_{\bs{k}}(\nu) &= \widetilde{F}^{t \up}_{\kk}(\nu)   
\end{align*}
We can, then, use decomposition~\eqref{eq: G dec} to write
\begin{align}
    F^{\rm m}_{\bs{k}}(\nu) &= \widetilde{F}^{\rm m}_{\bs{k}}(\nu) = \sum_l \frac{u^{12}_{l,\bs{k}}}{i\nu - E^l_{\bs{k}}}
\end{align}
for each Green's function component in~\eqref{eq: J in} and~\eqref{eq: J out}, making the Matsubara summation analytically doable. 

\end{appendix}

\bibliography{main.bib}

\end{document}